\documentclass[journal, twocolumn]{IEEEtran}
\makeatletter
\def\ps@headings{%
\def\@oddhead{\mbox{}\scriptsize\rightmark \hfil\thepage}%
\def\@evenhead{\scriptsize\thepage \hfil \leftmark\mbox{}}%
\def\@oddfoot{}%
\def\@evenfoot{}}
\makeatother
\IEEEoverridecommandlockouts

\usepackage{dblfloatfix}    
\usepackage{threeparttable}
\usepackage[dvipsnames]{xcolor}
\usepackage{calc}
\usepackage[space]{cite}
\usepackage{boxedminipage}
\usepackage{adjustbox}
\usepackage[latin9]{inputenc}
\usepackage{amsfonts}
\usepackage{diagbox}
\usepackage{graphicx}
\usepackage{amssymb}
\usepackage{standalone}
\usepackage{stmaryrd}
\usepackage{amsmath, amsthm}
\usepackage{autobreak}
\usepackage{verbatim}
\usepackage[noend]{algorithmic}
\usepackage[linesnumbered,ruled,vlined,boxed,commentsnumbered]{algorithm2e}

\usepackage{amsmath}
\usepackage{tikz}
\usetikzlibrary{patterns}
\usepackage{pgfplots}
\usepackage{pgfplotstable}
\usepackage{graphicx}
\usepackage{subfigure}
\usepackage[abs]{overpic}
\usepackage{multirow}
\usepackage{paralist}
\usepackage{wrapfig}
\usepackage{booktabs}
\usepackage{url}
\usepackage{float}
\usepackage{capt-of}
\usepackage[T1]{fontenc}
\usepackage{siunitx}
\usepackage{mathtools}

\usepackage{array}
\newcolumntype{P}[1]{>{\centering\arraybackslash}p{#1}}

\pgfplotsset{compat=1.15}
\newcommand{\Cross}{$\mathbin{\tikz [x=1.4ex,y=1.4ex,line width=.4ex, red] \draw (0,0) -- (1,1) (0,1) -- (1,0);}$}%
\newcommand{\remove}[1]{}

\usepackage{bm}
\usetikzlibrary{patterns}
\definecolor{mygreen}{RGB}{88,212,88}
\definecolor{mypink1}{rgb}{0.858, 0.188, 0.478}
\definecolor{myred}{RGB}{223,42,42}
\definecolor{myblue}{RGB}{30,144,255}
\definecolor{mygreen0}{RGB}{229,245,224}
\definecolor{mygreen1}{RGB}{161,217,155}
\definecolor{mygreen2}{RGB}{49,163,84}
\definecolor{myblue2}{RGB}{65,182,196}
\definecolor{myblue1}{RGB}{49,130,189}
\definecolor{mypink2}{RGB}{247,104,161}
\definecolor{mypink3}{cmyk}{0, 0.7808, 0.4429, 0.1412}
\definecolor{mygray}{gray}{0.6}
\usepackage[colorinlistoftodos]{todonotes}

\allowdisplaybreaks 
\theoremstyle{definition}

\usetikzlibrary{patterns}
\usetikzlibrary{shapes.geometric, arrows}
\tikzstyle{startstop} = [rectangle, rounded corners, minimum width = 1cm, minimum height=0.5cm, text centered, draw = black, fill = red!40]
\tikzstyle{decision} = [diamond, aspect = 3,text centered,draw=black,fill = green!30 ]
\tikzstyle{process} = [rectangle, minimum width=1.5cm, minimum height=0.5cm, text centered, draw=black, fill = yellow!50]
\tikzstyle{arrow} = [->,>=stealth]

\newcommand{\arcsinh}{\textrm{arcsinh}} 

\begin{document}
\title{Maximizing Revenue with Adaptive Modulation and Multiple FECs in Flexible Optical Networks}

\author{Cao~Chen,
        Fen~Zhou,~\IEEEmembership{Senior~Member,~IEEE,}, Massimo Tornatore,~\IEEEmembership{Senior~Member,~IEEE},
         Shilin~Xiao
\thanks{Cao Chen is with the State Key Laboratory of Advanced Optical Communication Systems and Networks, Shanghai Jiao Tong University, Shanghai, 200240, China (email: cao.chen@alumni.univ-avignon.fr). Cao Chen is also with the  CERI-LIA in University of Avignon, France.}

\thanks{Fen Zhou is with the CERI-SN, IMT Lille Douai, Institut Mines-T\'el\'ecom, University of Lille, Villeneuve-d'Ascq, 59650, France. He is also with the CERI-LIA, University of Avignon, France. (email: fen.zhou@imt-lille-douai.fr).}



\thanks{Massimo Tornatore is with the Department of Electronics, Information and Bioengineering in Politecnico di Milano, Italy (email: massimo.tornatore@polimi.it).}

\thanks{Shilin Xiao is with the State Key Laboratory of Advanced Optical Communication Systems and Networks, Shanghai Jiao Tong University, Shanghai, 200240, China (email: slxiao@sjtu.edu.cn).}

\thanks{A preliminary version of this work was presented as a short paper at IEEE HPCC 2019 \cite{ChZX19}.}
}

\maketitle
\begin{abstract}
Flexible optical networks (FONs) are being adopted to accommodate the increasingly heterogeneous traffic in today's Internet. However, in presence of high traffic load, not all offered traffic can be satisfied at all time. As carried traffic load brings revenues to operators, traffic blocking due to limited spectrum resource leads to revenue losses. In this study, given a set of traffic requests to be provisioned, we consider the problem of maximizing operator's revenue, subject to limited spectrum resource and physical layer impairments (PLIs), namely amplified spontaneous emission noise (ASE), self-channel interference (SCI), cross-channel interference (XCI), and node crosstalk. In FONs, adaptive modulation, multiple FEC, and the tuning of power spectrum density (PSD) can be effectively employed to mitigate the impact of PLIs. 
Hence, in our study, we propose a universal bandwidth-related impairment evaluation model based on channel bandwidth, which allows a performance analysis for different PSD, FEC and modulations.
Leveraging this PLI model and a piecewise linear fitting function, we succeed to formulate the revenue maximization problem as a mixed integer linear program. Then, to solve the problem on larger network instances, a fast two-phase heuristic algorithm is also proposed, which is shown to be near-optimal for revenue maximization. Through simulations, we demonstrate that using adaptive modulation enables to significantly increase revenues in the scenario of high signal-to-noise ratio (SNR), where the revenue can even be doubled for high traffic load, while using multiple FECs is more profitable for scenarios with  low SNR. 

\end{abstract}
\begin{keywords}
Flexible optical networks (FONs); revenue maximization; adaptive modulation; multiple forward-error correction (FEC);
\end{keywords}
\section{Introduction}

According to recent traffic reports, network traffic (fueled by successful network services like video on demand, file sharing, online gaming, and video conferencing) is still growing exponentially in today's Internet\cite{cisco2017}. This constant traffic growth can be accommodated by novel flexible optical networks (FONs) which can support large transmission capacity.
As busy hour traffic peaks are expected to increase almost 5 times between 2017 and 2022 (average Internet traffic will increase only 3.7 times), the problem of coping with sudden resource crunches will become even more a matter of concern in next years, especially during peak usage periods\cite{LTMM18,cisco2017,ZHTL19, LeZh15}. During resource crunch, given the limited spectrum resources in FONs, not all traffic requests can be fully satisfied and some traffic must be blocked. As carried traffic brings revenue to operators, resource crunch events can lead to significant revenue losses for operators. Hence, efficient provisioning strategies in optical networks are required to reduce blocking and maximize operators' revenue.

In FONs, an adequate amount of spectrum resources to establish a lightpath is required for each request.  
Since FONs can support variable routes, bandwidth, and modulation formats (MFs), the routing and wavelength assignment problem 
has evolved into the routing and spectrum assignment\cite{KlWa11,RTPG13, YADW17, MCKT09}. However, since the spectral efficiency granularity of $m$-ary quadrature amplitude modulation ($m$QAM) is coarse, the conventional resource provisioning cannot give full play to its advantages in collecting the services' revenue. Hence, to achieve even higher resource-allocation flexibility that granted by multiple MFs, tunability of forward error-corrections has been introduced to adjust the spectral efficiency\cite{SMCF15, AISB16,Bosc19,KPSS14}. It can be typically observed that overhead ratios range values from 7\% to 20\%. 
The combination of MF and FEC, referred to as \textit{transmission mode} in this paper, can provide more candidate choices in terms of spectral efficiency and transmission reach. 
Compared to the traditional approaches aiming at revenue improvement, such as using backup lightpaths for living traffic  \cite{VRPS11} or upgrading to multi-core fibers\cite{Koro12}, using MF and FEC is more efficient and faster. The traffic provisioning with multiple MFs and FECs maps into a problem  of routing, MF, FEC, and spectrum assignment (RMFSA) \cite{STCT19}. Although some researchers have proposed heuristic algorithms\cite{SMCF15,KPSS14, MCKT09}, like \textit{congestion-aware sequential loading} algorithm\cite{Savo13}, and \textit{adaptive FEC selection}\cite{LDSB14}, there is no complete mathematical model for the lightpath provisioning in FONs with both MF and FEC, which also accounts for physical layer impairments (PLIs) modeling.
While the optimal combination of MF and FEC has been investigated at the transmission layer in, \textit{e.g.}, \cite{KKMP17}, in this paper we investigate how the combination of MF and FEC can be used to maximize revenues through appropriate traffic provisioning strategies.




To support multiple MFs and FECs in FONs, traffic provisioning strategies must be cross-layer\cite{CDSK19}, \textit{i.e.}, they must be capable of taking in account physical layer aspects, to achieve efficient spectrum usage. The PLIs of a lightpath are influenced by the bandwidth, by power spectral densities (PSDs), by the route length, and by the number of crossed nodes. Due to the impact of all the parameters just mentioned, the quality of transmission (QoT) (\textit{e.g.}, expressed by ligthpath's Signal to Noise Ratio, SNR) may deteriorate and fall below acceptable threshold for correct signal reception after a long distance. Recent studies on the node crosstalk have considered the wavelength-related and frequency slot-related crosstalk component\cite{BDDM19,ChMV10}. However, current studies overestimate the PLIs with the assumption of full wavelength or full consecutive frequency slots for each lightpath. For example, the node crosstalk on provisioned bandwidth of 12.5 GHz~slot is larger than the actual value of the sub-channel bandwidth with 6.25~GHz\cite{KRSM15,MCDF17, SZZS18}.  To reduce the PLIs, a guard band (\textit{e.g}, 12.5 GHz) may be used, but incurring in inefficient usage of spectrum resources. Therefore, the PLI model that we propose is based on channel bandwidth, which means that the impairments of node crosstalk and fiber nonlinear interference are evaluated by the bandwidth rather than a wavelength or slot, which ensures that the PLIs are properly estimated and spectrum resource is effectively utilized.

The main novelty and contributions of this paper can be summarized as follows: 
\begin{enumerate}
    
\item We devise novel traffic provisioning strategies to maximize the total  revenue using different MF and FEC configurations. The lightpaths can adopt transmission mode with either higher spectral efficiency or longer transmission reach to guarantee the bit-rate under resource crunch. Compared to single MF or single FEC, the combination can provide just-enough spectral efficiency and transmission reach thus improve the traffic provisioning. By using a piece-wise linear fitting function to model the nonlinear interference and calculating the crosstalk of intermediate nodes, we  linearize the PLIs then model the studied problem as a mixed integer linear program (MILP). Our MILP model is based on flow rather than pre-calculated route.  Without using the candidate route, our method can get the optimal solution irrespective of the number of routes.

\item We propose a novel lighpath's PLI evaluation model based on the channel bandwidth that incorporates the impact of different PSDs, FECs, and MFs. By tracing the relationship between the PSD and SNR, we observe that using MFs enables to increase revenues with high SNR, while using multiple FECs is preferred for the scenarios of low SNR. Besides, compared to the wavelength-related or frequency slot-related method, the bandwidth-related method evaluates the PLIs by using channel bandwidth. To this end, the spectrum resources of fiber are assumed to be continuous rather than discrete frequency slot.


\item A fast and near-optimal heuristic algorithm is also proposed to solve the revenue maximization problem.
    
\end{enumerate}

The rest of this paper is organized as follows.
We 
describe our proposed PLI model in Sec. \ref{sec: PLIs}. The problem of traffic provisioning with adaptive MFs and multiple FECs is stated in Sec. \ref{sec: serviceprovisoning}. To solve it, we present a MILP model in Sec. \ref{sec: MILP} and a heuristic algorithm in Sec. \ref{sec: heu}. Illustrative numerical results are presented in Sec. \ref{sec: simulation}. Finally, Sec. \ref{sec: conclusions} concludes this paper.

\section{Physical Layer Model} \label{sec: PLIs}

In this section, we discuss our proposed PLI evalution model. We also include a  description of the signal impairments and of the optical transmission. 


\subsection{MF and FEC}




We denote as transmission mode $\mathcal{C}$ the combination set of MF and FEC, $\mathcal{C}$=($\mathcal{M}$,$\mathcal{F}$). 
For an arbitrary transmission mode $c\in \mathcal{C}$, its spectral efficiency is 
\begin{align}
SE(c)= SE(m,f) =  m/(1+OH_f) 
\end{align}
where $m$ is the theoretically maximum spectral efficiency of the MF, and $OH_f$ is the FEC overhead ($\times$100\%)\cite{AISB16,KKMP17}. The spectral efficiency and SNR threshold of several transmission modes are shown in Fig. \ref{fig: node_se(SNR)}. In particular, we assume four available  polarization-multiplexing MFs (PM-BPSK, PM-QPSK, PM-8QAM, and PM-16QAM) and six FEC OHs (1\%, 7\%, 10\%, 20\%, 30\%, and 50\%).
For example,  the spectral efficiency of PM-16QAM with FEC OHs 10\% is 4/(1+10\%)=3.63 bit/s/Hz, while the SNR threshold is 15.7 dB. 
To satisfy QoT, the SNR should be over the threshold for each transmission mode. Next, we present the PLI model that impacts SNR.
\begin{figure}[!htbp]
\flushleft     
\subfigure{
\begin{tikzpicture}
\begin{axis}[
ylabel shift = -2 pt,
        width=8.5cm,
        height=5cm,
        ymin=0, ymax=10,
	    xmin=0, xmax=25,
	    ylabel={SE [bit/s/Hz]},
        xlabel={SNR [dB]},
        mark options = {solid},
        extra y ticks={},
        extra y tick labels={},
        extra y tick style={grid=major},
        legend pos= south east,
        legend style={font=\scriptsize},
        legend columns = 1,
        xmajorgrids,
        ymajorgrids,
        grid style=dashed,
        legend entries={BPSK,
        QPSK,
        8QAM,
        16QAM,
        Shannon limit},
       label style={font=\small},
       tick label style={font=\small},
        ]
        \addlegendimage{very thick,draw=black, mark size=2pt}
\addlegendimage{very thick, mark size=2pt, mark=pentagon*, white!20!red, mark options={draw=black,fill=white,thick}, fill=white}
\addlegendimage{very thick, mark size=2pt, mark=diamond*, white!20!blue,mark options={draw=black,fill=white,thick}, fill=white}
\addlegendimage{very thick, mark size=2pt, mark=*, white!20!cyan,mark options={draw=black,fill=white,thick}, fill=white}
\addlegendimage{very thick, mark size=2pt, mark=triangle*, white!20!green, mark options={draw=black,fill=white,thick}, fill=white}

       \pgfplotstableread{TON/SNR_SE.txt}\loadedtable
       \addplot[white!20!black,very thick] table [x index=8, y index=9] {\loadedtable};
       \addplot[white!20!red,very thick] table [x index=6, y index=7] {\loadedtable};
       \addplot[white!20!blue,very thick] table [x index=4, y index=5]       {\loadedtable};     \addplot[white!20!cyan,very thick] table [x index=2, y index=3]       {\loadedtable};     \addplot[white!20!green,very thick] table [x index=0, y index=1] {\loadedtable};
       \addplot[mark=triangle*,fill=white, draw=black, mark size=2pt, only marks, thick] table [x index=10, y index=11] {\loadedtable};
        \addplot[mark=*,fill=white, draw = black, mark size=2pt,thick,only marks] table [x index=12, y index=13] {\loadedtable};
        \addplot[mark=diamond*,fill=white, draw=black,mark size=2pt,thick,only marks] table [x index=14, y index=15] {\loadedtable};
        \addplot[mark=pentagon*,fill=white, draw=black,mark size=2pt,thick,only marks] table [x index=16, y index=17] {\loadedtable};
\legend{Shannon limit, PM-16QAM, PM-8QAM, PM-QPSK, PM-BPSK}; 

\node[above right] at (15.7,8) {\tiny{PM-16QAM$_{1\%}$: (SNR$^\textrm{th}$, SE)}};
\node[above right] at (13.6,6) {\tiny{PM-8QAM$_{1\%}$: (SNR$^\textrm{th}$, SE)}};
\node[above right] at (6,1.98)  {\tiny{PM-BPSK$_{1\%}$: (SNR$^\textrm{th}$, SE)}};

      \end{axis}
\end{tikzpicture}
}
\subfigure
{\label{tab: modulation}

\begin{tabular}{p{2.0cm}p{0.8cm}p{0.9cm}p{0.9cm}p{0.9cm}p{0.9cm}p{0.9cm}p{0.9cm}p{1.2cm}}
\toprule
\multicolumn{1}{c}{\multirow{2}{*}{MF}} & \multicolumn{1}{c}{\multirow{2}{*}{$m$}} & \multicolumn{6}{c}{FEC}                              \\
\cmidrule(lr){3-8}
\multicolumn{1}{c}{}                    & \multicolumn{1}{c}{}                   & \multicolumn{1}{c}{OH=1\%} & \multicolumn{1}{c}{OH=7\%} & \multicolumn{1}{c}{OH=10\%} & \multicolumn{1}{c}{OH=20\%} & \multicolumn{1}{c}{OH=30\%} & \multicolumn{1}{c}{OH=50\%} \\

\cmidrule(lr){3-3}\cmidrule(lr){4-4}
\cmidrule(lr){5-5}\cmidrule(lr){6-6}
\cmidrule(lr){7-7}\cmidrule(lr){8-8}
  
  \midrule
  PM-BPSK & 2 & 6.02 & 3.56 & 2.95 & 1.6 & 0.67 & - \\
  PM-QPSK & 4 & 9.03 & 6.52 & 5.93 & 4.58 & 3.65 & 2.3  \\
  PM-8QAM & 6 & 13.6 & 10.98 & 10.31 & 8.85 & 7.81 & 6.26 \\
  PM-16QAM& 8 & 15.7 & 13.1 & 12.25 & 10.78 & 9.65 & 7.98\\
\bottomrule
\end{tabular}

}
\caption{The maximum achievable spectral efficiency of different transmission modes with different SNRs\cite{Yueqian11,AISB16}. With a target pre-FEC BER of 10$^{-4}$, the SNR threshold for each marker is illustrated in the table below.
}
\label{fig: node_se(SNR)}
\end{figure}

\subsection{PLIs model}
When an optical signal propagates, it suffers diverse forms of PLIs, including white Gaussian noise amplified spontaneous emission (ASE), self-channel interference (SCI), and cross-channel interference (XCI) \cite{Pogg12,JoAg14,YADW17}. 
Both SCI and XCI interference are caused by the Kerr effect of fiber, which can be estimated as additive white Gaussian noise by the GN model\cite{Pogg12}. When a signal traverses an optical cross-connect (OXC), the node crosstalk from signal adding or dropping (AD) at node must be also considered\cite{CDSK19, BDDM19}.

The PLI related parameters are given in Table \ref{tab: pli_notations}. We note $c_i$ as the transmission mode used by request $i$. $\textrm{SNR}_i$ denotes its received signal-to-noise ratio, and $\textrm{SNR}_{c_i}^{\textrm{th}}$ is the SNR threshold for $c_i$. By summing up all noise contributions due to PLIs, a request $i$ can be served if it satisfies the QoT constraint as in Eq. (\ref{eq: OSNR origin}).
\begin{align}
\label{eq: OSNR origin}
& \textrm{SNR}_i= \frac{G_i}{G_i^{\textrm{ASE}}+G_i^{\textrm{SCI}}+G_i^{\textrm{XCI}}+G_i^{\textrm{AD}}}  \geq \textrm{SNR}^{\textrm{th}}_{c_i}
\end{align}

For ease of the MILP modeling, we can also express Eq. (\ref{eq: OSNR origin}) in its  reciprocal  form,
\begin{align}
\label{eq: OSNR newform}
&\left\{
\begin{aligned}
\frac{1}{\textrm{SNR}_i}&=t_i^{\textrm{ASE}}+ t_i^{\textrm{SCI}}+ t_i^{\textrm{XCI}}+ t_i^{\textrm{AD}} \leq \frac{1}{\textrm{SNR}^\textrm{th}_{c_i}} \\
t_i^{\textrm{ASE}}&=G_i^{\textrm{ASE}}/G_i \\
t_i^{\textrm{SCI}}&=G_i^{\textrm{SCI}}/G_i \\
t_i^{\textrm{XCI}}&=G_i^{\textrm{XCI}}/G_i \\
t_i^{\textrm{AD}}&=G_i^{\textrm{AD}}/G_i
\end{aligned}
\right.
\end{align}
where $t_i^{\textrm{ASE}}$, $t_i^{\textrm{SCI}}$, $t_i^{\textrm{XCI}}$, and $t_i^{\textrm{AD}}$ are the noise to signal ratios of  ASE, SCI, XCI, and AD node crosstalk, respectively. The noise to signal ratios can be regarded as the amount of PLIs of ASE, SCI, XCI, and node crosstalk. In the following, we will explain in detail the computation of various  impairments.

\begin{table}[!htbp]
    \centering
    \caption{Parameters for PLIs}
    \label{tab: pli_notations}
    \scalebox{1}{
    \begin{tabular}{p{1cm}p{7.2cm}}
    \toprule
  \multicolumn{2}{c}{\textbf{Parameters and description}}\\
  \midrule
 $\alpha $ & Power attenuation ratio of fiber, 0.2 dB/km.\\
$\beta_2$ & Second order dispersion of 1550nm wavelength, -21.7 ps$^2$/km. \\
$\gamma$ & Non-linear coefficient, 1.3~(W$\cdot$Km)$^{-1}$. \\
$h$ & Planck's constant.\\
$\nu$ & Frequency of optical signal, 192.5~THz.\\
$\mu$ & $\frac{3\gamma^2}{2\pi \alpha \beta_2}$. \\
$\rho$ & $\pi^2 \beta_2/\alpha$.\\
$\epsilon_{\textrm{X}}$ & OXC port leakage ratio, -25~dB.\\
$n_{\textrm{sp}}$ & Noise factor of optical ampilier, 7dB.\\ 
$G_{i}$ & Power spectral density of request $i$.\\ 
$G_{i}^{\textrm{ASE}}$ & PSD of ASE.\\
$G_{i}^{\textrm{SCI}}$ & PSD of SCI.\\
$G_{i}^{\textrm{XCI}}$ & PSD of XCI.\\
$G_{ijv}^{\textrm{AD}}$ & PSD of AD node crosstalk from $j$ to $i$.\\
$G_{i}^{\textrm{AD}}$  & PSD of AD, $G_{i}^{\textrm{AD}}=\sum_{jv} G_{ijv}^{\textrm{AD}}$.\\
$\textrm{SNR}_{c}^{\textrm{th}}$ &  SNR threshold of transmission mode $c$.\\
$L_{\textrm{span}}$ & Span length, 100 km/span.\\
\bottomrule
\end{tabular}
}
\end{table}

\subsubsection{Impairments along fibers (ASE, SCI, and XCI)}
ASE noise is a white noise, whose intensity is proportional to the number of fiber spans and channel bandwidth. Its PSD can be expressed by
\begin{align}\label{eq: GiASE}
G_i^{\textrm{ASE}} &= N_{\textrm{span}} (e^{\alpha L_{\textrm{span}} } -1) n_{\textrm{sp}}h\nu 
\end{align}
where 
$N_{\textrm{span}}$ is the number of spans (see Table \ref{tab: pli_notations} for the other parameters).

In GN model\cite{Pogg12}, both SCI and XCI are regarded as white noise, whose PSD is related to the light power, bandwidth and center frequency. Eqs. (\ref{eq: GiSCI}) and  (\ref{eq: GiXCI}) can be used to calculate the PSD of SCI and XCI\cite{PBCC14,JoAg14} (note that the calculation has been validated for bandwidth $\Delta f_i$ bigger than 28 GHz\cite{JoAg14}.

\begin{align}\label{eq: GiSCI}
G_i^{\textrm{SCI}} &= N_{\textrm{span}} \mu G_i^3 \arcsinh(\rho \Delta f_i^2)\\
\label{eq: GiXCI}
G_i^{\textrm{XCI}} &= \sum_j  N_{\textrm{span},ij} \mu G_i G_j^2 \ln \left( \frac{\left|f_i-f_j\right|+\Delta f_j/2}{\left|f_i-f_j\right|-\Delta f_j/2}\right)
\end{align}
where $\Delta f_i$ is the bandwidth of request $i$ (unit: GHz), and $f_i$ is the relative carrier center frequency (unit: GHz). 

\subsubsection{Impairments at nodes }

\begin{figure}[!htbp]
    \centering
    \subfigure[]{
    \label{fig: 9node}
    \scalebox{0.9}{
    \includegraphics[width = 8cm]{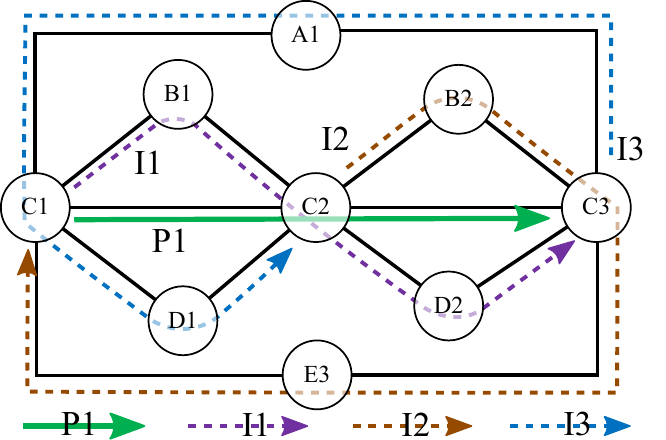}
    }
    }
    \subfigure[]{
    \label{fig: BS}
    \includegraphics[width = 8cm]{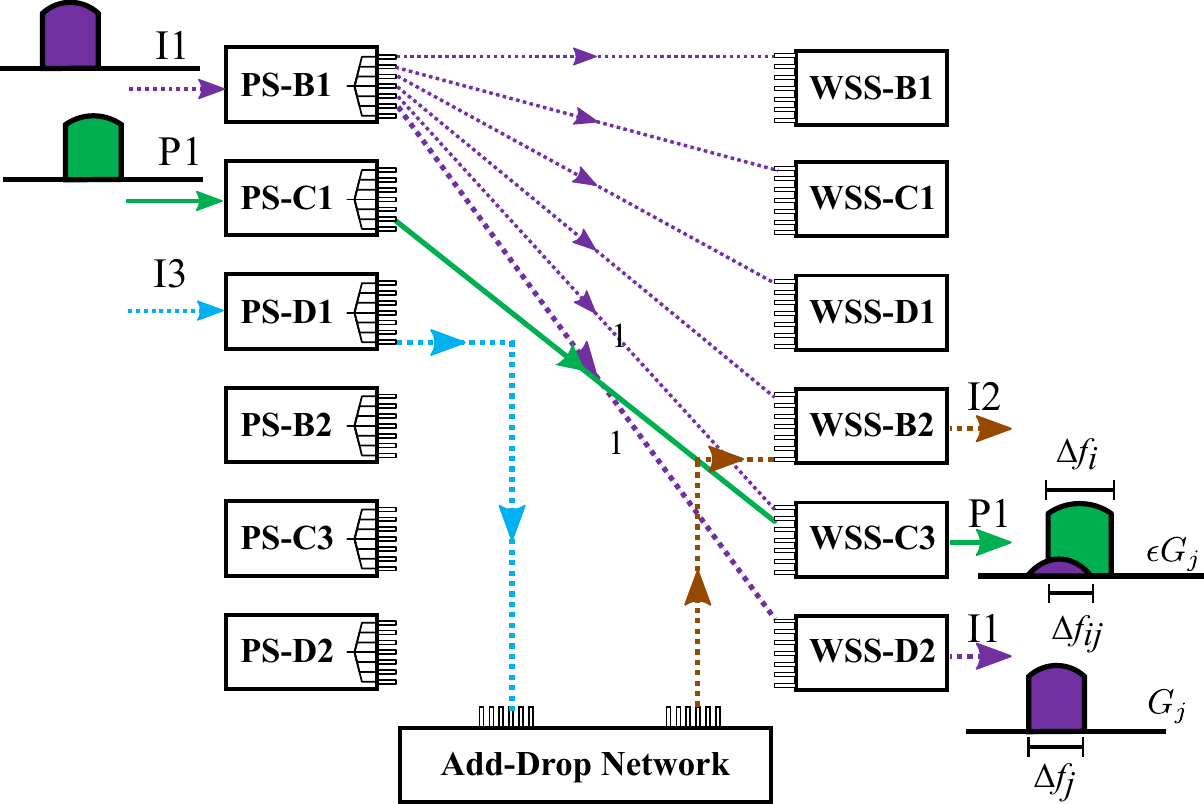}
    }
    \caption{(a) The 9-node network used to  illustrate different forms of AD node crosstalk, (b) Illustration of node crosstalk at node C2 considering a B\&S architecture\cite{ZSPF16}. PS: power splitter (PS). }\label{fig: node_xt}
\end{figure}

Impairments at nodes come from filtering effects and node crosstalk. In-band  crosstalk  is considered  in  this  paper, \textit{i.e.} the  lightpath  experiences  node crosstalk if  it  is  exposed  to  the  other  lightpaths  with  the overlapping  bandwidth.  As an example, we use 9-node network in Fig. \ref{fig: 9node} to illustrate the different node crosstalk components. The primary signal P1 is added at node C1, passes through node C2, and is dropped at node C3. Other three crosstalk signals I1, I2, and I3, are also depicted. At each of its nodes(C1, C2 and C3), P1 will experience all forms of crosstalk, the primary signal will experience all forms of crosstalk, \textit{i.e.} adding, passing through, and dropping.

To analyse the node crosstalk, we assume a broadcast-and-select (B\&S) OXC architecture for intermediate node C2\cite{ZSPF16}, as illustrated in Fig. \ref{fig: BS}. It consists of passive optical splitters (PSs) with 1$\times$N ports that broadcast signal copies to the common port at local add/drop side and wavelength
selective switches (WSSs) facing different output ports and collecting the signals from add ports. 
A node crosstalk from signal I1 to P1 arises on WSS-C3 because the broadcast signal I1 is leaked into WSS-C3.
We can also observe that the crosstalk signal I3 can leak into P1 on WSS-C3 due to the broadcast function of PS-D1. In addition, we also observe other forms of interference at nodes C1 and C3, which are not shown in Fig. \ref{fig: BS}. When primary signal P1 is added at node C1, it experiences the dropping crosstalk of I2 and the passing-through crosstalk of I3. 
When the primary signal is dropped at node C3, there is no crosstalk, because P1 is dropped locally. In short, the AD node crosstalk exists if primary signal is added at or passing through the node, and crosstalk signal is passing through or dropped at that node.


Different from the approach of \cite{BDDM19} that only supports the node crosstalk by fixed frequency slot, we propose to improve it by adopting the channel bandwidth, which supports arbitrary  bandwidth. Thus, assuming the overlapping bandwidth between the primary signal $\Delta f_{i}$ and the interfering signal $\Delta f_{j}$ is $\Delta f_{ij}$, we give the amount of node crosstalk as follows,
\begin{align}
\label{eq: AD}
G_{ijv}^{\textrm{AD}} = & \epsilon_{\textrm{X}} \Delta f_{ij} G_{j}
\end{align}
where $G_j$ is the PSD of other interfering signal ($j$ = I1 in the example), $v$ is the node C2, and $\Delta f_{ij} = \left| \frac{\Delta f_i + \Delta f_j}{2} - |f_i -f_j| \right|$. In addition, P1 also experiences the crosstalk of I2 at node C1 and the crosstalk of I3 at node C2.


\section{Traffic Provisioning using MF and FEC in Flexible Optical Networks}\label{sec: serviceprovisoning}

We denote a FON by a graph $\mathrm{G}(V,E)$. Each node $v\in V$ represents an OXC. A link $e\in E$ represents two fibers $uv$ and $vu$ ($u,v\in V$) that carry traffic in opposite directions.  Each request $i$ is characterized by its source node $s_i$, destination node $d_i$, bit-rate $r_i$ (Gbps), PSD $G_i$ ($G_i=G$), and revenue level $\eta_i$. The consumed spectrum bandwidth counts $\Delta f_i=r_i/\text{SE}(c_i)$ , where SE($c_i$) is the spectral efficiency of transmission mode $c_i$. Normally, the revenue $\eta_i$ is determined by the operator's preference or the importance, such as  (\textit{i}) time of day, (\textit{ii}) duration, (\textit{iii}) location, (\textit{iv}) distance, (\textit{v}) connection speed, and (\textit{vi}) service type \cite{LTMM18}. A random service type parameter is adopted in this paper. Assuming a set of requests in demand $D$, the total network revenue is the sum of accepted requests' revenue. 
Available spectrum resource of each fiber is limited to $F$ ($F\in \mathbb{R}^+$).

To serve a request, a lightpath should be established on a continuous and contiguous spectrum interval. We indicate the continuous spectrum interval as $[b_i,e_i]$, where $b_i$ and $e_i$ are the beginning and end of the spectrum interval of request $i$, respectively. To satisfy the spectrum continuity and contiguity constraints, the spectrum interval $[b_i,e_i]$ must be the same on all traversed links, and can not overlap with other lightpath. Due to limited spectrum resources, not all lightpaths can be provisioned. Therefore, our objective is to maximize the total revenue by optimizing  spectrum resource allocation. We use variable $B_i$ to indicate whether request $i$ is served ($B_i=1$ if it is accepted, 0 otherwise), then the objective function can be expressed by 
\begin{equation}
\label{eq: obj}
    \sum_{i\in D} \eta_i B_i 
\end{equation}

\begin{figure}[!htbp]
\centering     
\subfigure[6-node network]{
\includegraphics[width = 6.5cm]{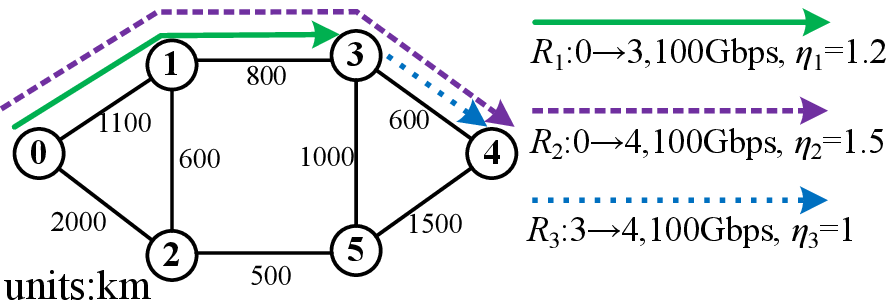}
}
\subfigure[Single-FEC]{

\definecolor{zzttqq}{rgb}{0.4,0.1,0.6}
\definecolor{ududff}{rgb}{0,0.69,0.32}
\definecolor{blue}{rgb}{0,0,1}

\begin{tikzpicture}[line cap=round,line join=round,>=triangle 45]
\begin{axis}[
width=8cm,
height=6.5cm,
axis lines=middle,
ymajorgrids=true,
xmajorgrids=true,
ticks=none,
xmin=0, xmax=3.05,
ymin=0, ymax=110,
xtick={0,1,2,3},
ytick={0,12.5,...,100},
axis line style = thick,
grid style =dashed,
]

\draw[rounded corners=5pt, fill = white, fill opacity=0.4, color=zzttqq, loosely dashed, line cap=rect, line width=2pt,] (0, 46.5) rectangle (1, 100) node [pos=.5] {};
\draw[rounded corners=5pt, fill = white,  fill opacity=0.4, color=zzttqq, loosely dashed, line cap=rect,  line width=2pt,] (1, 46.5) rectangle (2, 100) node [pos=.5] {};
\draw[rounded corners=5pt, loosely dashed, line cap=rect,  fill=zzttqq,fill opacity=0.4, color=zzttqq,line width=2pt,] (2, 46.5) rectangle (3, 100) node [pos=.5] {};
\draw[rounded corners=5pt, loosely dotted, line cap=butt, fill=blue,fill opacity=0.4, color=blue,line width=2pt,] (2, 0) rectangle (3, 26.7) node [pos=.5] {};

\begin{scriptsize}

\draw[color=black,right] (0,104.5) node {\Large Spectrum[GHz]};

\draw[color=black] (0.5,75) node {$BPSK_{7\%}$};
\draw[color=black] (1.5,75) node {$BPSK_{7\%}$};
\draw[color=black] (2.5,75) node {$BPSK_{7\%}$};
\draw[color=black] (2.5,15) node {$QPSK_{7\%}$};

\draw[color=black] (1.0,25) node {$BPSK_{7\%}$};
\end{scriptsize}
\end{axis}

\draw[color=black] (1,0) node[anchor=north] {link 0-1};
\draw[color=black] (3,0) node[anchor=north] {link 1-3};
\draw[color=black] (5,0) node[anchor=north] {link 3-4};

\draw[color=black,left] (0,0) node {0};
\draw[color=black,left] (0,1.2) node {26.7};
\draw[color=black,left] (0,2.0) node {46.5};
\draw[color=black,left] (0,4.5) node {100};

\draw[rounded corners=5pt, fill=ududff, fill opacity=0.1, color=ududff,line width=2pt] (0, -0.5) rectangle (4.2, 2.05) node [pos=.5] {};
\draw[color=black,left] (2.4,-0.3) node {\Cross};

\end{tikzpicture}
}
\subfigure[Multiple-FEC]{\definecolor{zzttqq}{rgb}{0.4,0.1,0.6}
\definecolor{ududff}{rgb}{0,0.69,0.32}
\definecolor{blue}{rgb}{0,0,1}

\begin{tikzpicture}[line cap=round,line join=round,>=triangle 45]
\begin{axis}[
width=8cm,
height=6.5cm,
axis lines=middle,
ymajorgrids=true,
xmajorgrids=true,
ticks=none,
   grid style=dashed,
xmin=0, xmax=3.05,
ymin=0, ymax=110,
xtick={0,1,2,3},
ytick={0,12.5,...,100},
axis line style = thick,
]

\draw[rounded corners=5pt, loosely dashed, line cap=rect, fill=zzttqq,fill opacity=0.4, color=zzttqq,line width=2pt,] (0, 70) rectangle (1, 100) node [pos=.5] {};
\draw[rounded corners=5pt, loosely dashed, line cap=rect, fill=zzttqq,fill opacity=0.4, color=zzttqq,line width=2pt,] (1, 70) rectangle (2, 100) node [pos=.5] {};
\draw[rounded corners=5pt, loosely dashed, line cap=rect, fill=zzttqq,fill opacity=0.4, color=zzttqq,line width=2pt,] (2, 70) rectangle (3, 100) node [pos=.5] {};
\draw[rounded corners=5pt, loosely dotted, line cap=butt, fill=blue,fill opacity=0.4, color=blue,line width=2pt,] (2, 0) rectangle (3, 30) node [pos=.5] {};

\draw[rounded corners=5pt, fill=ududff,fill opacity=0.4, color=ududff, line width=2pt,] (0, 35) rectangle (1, 65) node [pos=.5] {};
\draw[rounded corners=5pt, fill=ududff,fill opacity=0.4, color=ududff,line width=2pt,] (1, 35) rectangle (2, 65) node [pos=.5] {};

\begin{scriptsize}
\draw[color=black] (0.5,85) node {$QPSK_{20\%}$};
\draw[color=black] (1.5,85) node {$QPSK_{20\%}$};
\draw[color=black] (2.5,85) node {$QPSK_{20\%}$};
\draw[color=black] (2.5,15) node {$QPSK_{20\%}$};
\draw[color=black] (0.5,50) node {$QPSK_{20\%}$};
\draw[color=black] (1.5,50) node {$QPSK_{20\%}$};

\draw[color=black,right] (0,105) node {\Large Spectrum[GHz]};


\end{scriptsize}
\end{axis}

\draw[color=black] (1,0) node[anchor=north] {link 0-1};
\draw[color=black] (3,0) node[anchor=north] {link 1-3};
\draw[color=black] (5,0) node[anchor=north] {link 3-4};

\draw[color=black,left] (0,0) node {0};
\draw[color=black,left] (0,1.3) node {30};
\draw[color=black,left] (0,1.6) node {35};
\draw[color=black,left] (0,2.9) node {65};
\draw[color=black,left] (0,3.2) node {70};
\draw[color=black,left] (0,4.4) node {100};
\end{tikzpicture}
}
\subfigure
{
\scalebox{0.92}{
\begin{threeparttable}
\scalebox{1}{\begin{tabular}{@{}p{2cm}P{3cm}P{3cm}@{}}
\toprule
Request  & Single-FEC\tnote{\textit{i}} & Multiple-FEC\tnote{\textit{ii}}       \\ \midrule
\begin{tabular}[c]{@{}l@{}} 
R1: 0-1-3 \end{tabular} & blocked & $\textrm{QPSK}_{{20\%}}$ \\
\begin{tabular}[c]{@{}l@{}} R2: 0-1-3-4 \end{tabular} &  $\textrm{BPSK}_{{7\%}}$  & $\textrm{QPSK}_{{20\%}}$\\
\begin{tabular}[c]{@{}l@{}} 
R3: 3-4 \end{tabular} & $\textrm{QPSK}_{{7\%}}$ & $\textrm{QPSK}_{{20\%}}$ 
\\ \midrule
Revenue & 2.5  & 3.7 \\ \bottomrule
\end{tabular}
}
\begin{tablenotes}
        \footnotesize
        \item[\textit{i}] Single-FEC $\mathcal{C}$: \{$\textrm{BPSK}_{{7\%}}$,$\textrm{QPSK}_{{7\%}}$\} 
        \item[\textit{ii}] Multiple-FEC $\mathcal{C}$: \{$\textrm{QPSK}_{{7\%}}$,$\textrm{QPSK}_{{20\%}}$\}
 \end{tablenotes}
 \end{threeparttable}
}
}

\caption{Traffic provisioning example with single-FEC and multiple-FEC configurations in 6-node network.}\label{fig: instance}
\end{figure}

We give an example of revenue difference for different traffic provisioning using transmission mode configurations, \textit{i.e.} single-FEC and multiple-FEC in Fig.~\ref{fig: instance}. Each link  has limited spectrum of 100~GHz. The number near each link denotes the length. Three requests R1, R2, and R3 are labeled with source, destination, bit-rate, and revenue. The PSDs for all lightpaths are -11 dBm/GHz. 

In single-FEC configuration, FEC OH is fixed at 7\%, hence R2 uses BPSK$_{7\%}$ on the channel [46.5, 100], and R3 uses QPSK$_{7\%}$ on channel [0, 26.7]. However, the remaining spectrum [0, 46.5] on link 0-1 is not enough to support R1 with BPSK$_{7\%}$. Otherwise, if QPSK$_{7\%}$ had been chosen, the QoT of R1 would not have been satisfied because of the fiber nonlinear interference from R2 to R1. Therefore, only two requests can be accepted with single-FEC configuration, leading to a revenue of 2.5.

In multiple-FEC configuration, two FEC OHs can be used. The requests R1, R2, and R3 are served with  $\textrm{QPSK}_{{20\%}}$, as the SNRs  with $\textrm{QPSK}_{{20\%}}$ are 6.0~dB, 4.78~dB, and 10.8~dB, respectively, according to Eqs. (\ref{eq: OSNR origin}), (\ref{eq: GiASE}), (\ref{eq: GiSCI}), and (\ref{eq: GiXCI}), which are all over the threshold 4.58~dB. Hence, for multiple-FEC configuration, the revenue is 3.7.

From the example, we see that the revenues can be improved by using multiple-FEC configuration. The traffic provisioning is composed by routing, MF, FEC, and spectrum assignment. 

\section{MILP Formulation}\label{sec: MILP}

In this section, we formulate the revenue maximization problem as a MILP, named as RMAX. The parameters and variables of the MILP are summarized in Table \ref{tab: ILPpara}. 

\begin{table}[!htbp]
\caption{Parameters And Variables in RMAX}
\label{tab: ILPpara}
\scalebox{0.92}{
\begin{tabular}{p{1.8cm}p{6.9cm}}
  \toprule
  \multicolumn{2}{c}{\textbf{Network Sets and Parameters}}\\
  \midrule
$V,E$  & Node set and link set of the FON G.\\
$uv \in E $ & A link from node $u$ to $v$.\\
$s_i, d_i\in V$ & Source and destination node of request $i$.\\
$L_{uv}$ & Number of spans on link $uv$.\\
$F \in \mathbb{R}_{\geq 0}$ & Available spectrum resources of an optical fiber.\\
$N(v)$ & Adjacent node set of $v$ in $\textrm{G}$.\\
$D$ & Traffic demand matrix.\\
$i,j\in D$ & Any two requests $i$ and $j$ in traffic demand matrix $D$.\\
$r_i$  & Required bit-rate (Gbps) of request $i$.\\
$\eta_i$  & Revenue of request $i$.\\
$c\in \mathcal{C}$ & Transmission mode in candidate transmission mode set $\mathcal{C}$. \\
$\theta $ & A large constant.\\
$\epsilon_1$ & A factor balancing the importance between revenue and PLIs.\\
$o_k^1,o_k^0$ & Coefficients of piece-wise linear fitting function for fitting the XCI, $k\in\{1,2,...,Q\}$, where $Q$ is the number of segments.\\
\midrule
  \multicolumn{2}{c}{\textbf{Variables in RMAX}}\\
  \midrule
$B_{i} \in \{0,1\}$  & Equals 1 if request $i$ is accepted, 0 otherwise.\\ 
$q_{v}^i \in \{0,1\}$  & Equals 1 if request $i$ goes into node $v$, 0 otherwise.\\ 
$p_{v}^i \in \{0,1\}$  & Equals 1 if request $i$ goes out of node $v$, 0 otherwise.\\
$x_{uv}^{i} \in \{0,1\}$ & Equals 1 if request $i$ uses link $uv$, 0 otherwise. \\
$x_{uv,c}^{i} \in \{0,1\}$ & Equals 1 if request $i$ uses link $uv$ and transmission mode $c$, 0 otherwise.\\ 
$m_{c}^{i} \in \{0,1\} $ & Equals 1 if request $i$ uses transmission mode $c$, 0 otherwise.\\
$f_i \in [0,F] $ & Center frequency of request $i$.\\
$f_{ij} \in [0,F] $ & Center frequency difference  between requests $i$ and $j$.\\
$\Delta f_{i}\in[0,F]$ & Bandwidth of request $i$.\\
$\Delta f_{ij}\in [0,F]$ & Overlapping bandwidth between $i$ and $j$.\\
$f_{ij}^X\in [0,F]$ & Auxiliary variable of overlapping bandwidth $\Delta f_{ij}$.\\
$w_{ij} \in \{0,1\}$ & Equals 1 if $f_i$ is greater than $f_j$, 0 otherwise.\\ 
$t^{\textrm{ASE}}_i \in \mathbb{R}_{\geq 0}$ & PLI of ASE noise of request $i$, $G_i^\textrm{ASE}/G_i$.\\
$t^{\textrm{SCI}}_i \in \mathbb{R}_{\geq 0}$ & PLI of SCI of request $i$,  $G_i^\textrm{SCI}/G_i$.\\
$t_{ij,u}^{\textrm{XCI}} \in \mathbb{R}_{\geq 0}$ & Accumulated PLI of XCI of request $i$ from source node $s_i$ to $u$ that is generated by request $j$.\\
$t_{ij,v}^{\textrm{AD}} \in \mathbb{R}_{\geq 0}$ & Accumulated PLI of AD node crosstalk of request $i$ from source node $s_i$ to node $v$ that is generated by request $j$.\\
$h_{c}^{ij} $ & Piece-wise linear fitting term for XCI from $j$ to $i$ if $j$ takes transmission mode $c$. \\
$t^{\textrm{PLI}}_i \in \mathbb{R}_{\geq 0}$ & Total PLIs of request $i$. \\ 
\bottomrule
\end{tabular}
}
\end{table}

To provision the lightpaths, both \textit{network flow constraints} and \textit{spectrum assignment constraints} must be taken into account. Besides, \textit{SNR constraints} are also considered to ensure lightpaths' QoT. Thus, the problem is modeled as follows,

\begin{equation*}
\label{eqn: ILPobjective}
\max \quad  \sum_{i\in D} ( \eta_i B_i - \epsilon_1  t^{\textrm{PLI}}_i ) \quad \boldsymbol{\mathrm{(RMAX)}}
\end{equation*}
\textbf{\textit{\quad \quad \qquad s.t.}} Constraints (\ref{eq: netandroute})-(\ref{eqn: conPLI}).

\vspace{0.1em}

The main objective of this MILP is to maximize the total revenue and the second objective is to minimize the total PLIs for all requests. The second objective can reduce the PLIs and improve the SNR margin of network, which is regarded as an indirect way that can be used to guarantee the  revenue performance\cite{Yvan17}. The weighted factor $\epsilon_1$ is used to balance the importance between revenue and PLIs. For the sake of readability, we use $\forall i, \forall v, \forall uv,\forall c$ to denote $\forall i\in D, \forall v\in V, \forall uv\in E,\forall c\in \mathcal{C}$ in the following text.

\subsubsection{Network flow constraints} \noindent

\begin{subequations}
\label{eq: netandroute}
\resizebox{0.95\linewidth}{!}{
  \begin{minipage}{1.00\linewidth}
\begin{align}
\label{eq: nodemark1}
& q_{v}^i =\sum_{u\in N(v)}{x_{uv}^{i}}&&\forall i,\forall v\\
\label{eq: nodemark2}
& p_{v}^i = \sum_{u\in N(v)}{x_{vu}^{i}}&&\forall i,\forall v\\
\label{eq: flow-reservation}
&p_{v}^i - q_{v}^i=
\left\{
       \begin{aligned}
       & B_i,  && v=s_i \\
       & -B_i, && v=d_i \\
       & 0, && \textrm{others}. \\
       \end{aligned}
\right. &&\forall i, \forall v
\end{align}
  \end{minipage}
}
\end{subequations}

Constraints (\ref{eq: nodemark1}) and (\ref{eq: nodemark2}) determine the incoming and outgoing flow of request $i$ at node $v$. For any request $i$, the incoming degree $q_v^i$ counts 1 if request $i$ passes through or drops at node $v$. Also, for any request $i$, the outgoing degree $p_v^i$ counts 1 if request $i$ is added at or passes through node $v$. 
We will see that, with the help of   $q_v^i$ and $p_v^i$, we can calculate the node crosstalk and other PLIs node-by-node. Constraints (\ref{eq: flow-reservation}) are the flow conservation constraints. If a request $i$ gets accepted ($B_i$=1), there exists a lightpath from the source node $s_i$ to destination node $d_i$.


\subsubsection{Spectrum assignment constraints}\noindent

\begin{subequations}
\begin{align}
\label{eq: band1}
&\sum_{c\in \mathcal{C}}{m_{c}^{i}}=B_i &&\forall i\\
\label{eq: band2}
&\Delta f_i \geq \sum_{c \in \mathcal{C} }{\frac{r_i}{\mathrm{SE}(c)}m_{c}^{i}}  &&\forall i\\
\label{eq: wij}
	&w_{ij} + w_{ji}=1 && \forall i<j\\
\label{eqn: conspecCap}
&\left. \begin{aligned}
& f_{i}+\Delta f_i /2 \leq F  \\
& 0 \leq f_{i}-\Delta f_i /2 \\
  \end{aligned} 
  \right\} &&\forall i \\
\label{eqn: Fij}
&\left. \begin{aligned}
& f_{ij} \leq f_{i} - f_{j} + 2F(1-w_{ij})\\
      & f_{i} - f_{j} \leq f_{ij} \\
  \end{aligned} \right\}
&&\forall i<j\\
\label{eqn: spec1}
& \Delta f_{ij} \geq \min\left(\Delta f_i, \Delta f_j, f_{ij}^X \right) && \forall i<j \\
\label{eqn: spec2}
&\Delta f_{ij} \leq F  (2-x^{i}_{uv}-x^{j}_{uv}) && \forall uv, \forall i<j\\
\label{eqn: spec4}
&\left. \begin{aligned}
	&\Delta f_{ij}  = \Delta f_{ji}\\
	&f_{ij} = f_{ji}
  \end{aligned} \right\}
&& \forall  i<j
\end{align}
\end{subequations}

Constraints (\ref{eq: band1}) select one transmission mode for non-blocked request $i$ (if $B_i=1$). Constraints (\ref{eq: band2}) define the bandwidth of request $i$ by its bit-rate and the adopted transmission mode. Constraints (\ref{eq: wij}) assure that either $w_{ij}$ or $w_{ji}$ should be equal to 1. Constraints (\ref{eqn: conspecCap}) limit the fiber spectrum within $[0,F]$. Constraints (\ref{eqn: Fij}) calculate the frequency difference $f_{ij}$ between $i$ and $j$.
Constraints (\ref{eqn: spec1}) calculate the overlapping bandwidth $\Delta f_{ij}$. As it is not linear, we replace it by the following equations,
\resizebox{0.98\linewidth}{!}{
  \begin{minipage}{1.00\linewidth}
\begin{align*}
\Delta f_{ij} & \geq \min \left(\Delta f_i, \Delta f_j,  f_{ij}^X \right) \\ 
\Leftarrow &\left\{
\begin{aligned}
&\Delta f_i - F a^1_{ij} \leq \Delta f_{ij}\\
&\Delta f_j - F a^2_{ij} \leq \Delta f_{ij}\\
& f^X_{ij} - F a^3_{ij} \leq \Delta f_{ij}\\
& 0\leq  f^X_{ij}\\
& \frac{\Delta f_i + \Delta f_j}{2} -f_{ij} \leq f^X_{ij}\\
& f^X_{ij} = f^X_{ji}\\
& a^1_{ij}+a^2_{ij}+a^3_{ij}=2\\
& a^1_{ij},a^2_{ij},a^3_{ij}\in \{0,1\}
\end{aligned} \right.  \quad\quad\quad\quad  \forall i<j &&
\end{align*}
  \end{minipage}
}
\vspace{0.4em}

Constraints (\ref{eqn: spec2}) are spectrum non-overlapping constraints, indicating that when $i$ and $j$ share a common link, the overlapping bandwidth $\Delta f_{ij}$ on that link must be 0. Constraints (\ref{eqn: spec4}) guarantee that both variables $\Delta f_{ij}$ and $f_{ij}$ are symmetric.

\subsubsection{SNR constraints}\noindent


\begin{subequations}
\label{eqn: conPLI}
\begin{align}
\label{eqn: ASE}
&t^{\textrm{ASE}}_i =\sum_{uv} \frac{G_i^{\textrm{ASE}}}{G_i} L_{uv} x_{uv}^{i} \hspace{9em} \forall i\\
\label{eqn: SCI}
&t^{\textrm{SCI}}_i =\sum_{uv }\sum_{c} \mu G_i^2 L_{uv} \arcsinh\left(\rho \left(\frac{r_i}{\textrm{SE}(c)}\right)^2\right)  x^{i}_{uv,c} \hspace{1em} \forall i \\
\label{eq: flowlinktm}
&x^{i}_{uv,c} +1 \geq x^{i}_{uv} + m^{i}_{c} \hspace{6em} \forall i,\forall c, \forall uv\\
\label{eqn: XCIij}
\begin{split}
&t_{ij,v}^{\textrm{XCI}} - t_{ij,u}^{\textrm{XCI}} + \theta(2-x_{uv}^{i}-x_{uv,c}^{j}) \geq \mu G_j^2 h_c^{ij} L_{uv} \\
& \hspace{12em} \forall uv,\forall i\neq j,\forall c
\end{split}\\
\label{eqn: XCIi}
\begin{split}
&t_{ij,v}^{\textrm{XCI}} - t_{ij,u}^{\textrm{XCI}} + \theta(1-x_{uv}^{i}) \geq 0 \hspace{3em} \forall uv,\forall i\neq j
\end{split}\\
\label{eqn: conPLIAD}
\begin{split}
&t_{ij,v}^{\textrm{AD}} + \theta ( 3 - p^{i}_{v} - q^{j}_{v} - m^{j}_{c}) \geq \epsilon_{\textrm{X}} (\Delta f_{ij} G_j) / (r_i/\textrm{SE}(c) G_i) \\
& \hspace{8em} \forall c,\forall i\neq j, \forall v
\end{split}\\
\label{eqn: conPLI5}
& t_{i}^{\textrm{PLI}} \geq t^{\textrm{ASE}}_i+  t^{\textrm{SCI}}_i + \sum_{j \neq i}{t_{ij,d_i}^{\textrm{XCI}}} + \sum_{j \neq i} { {t_{ij,d_i}^{\textrm{AD}}}}  \hspace{2em} \forall i\\
\label{eqn: conPLI7}
& t_{i}^{\textrm{PLI}} \leq \sum_{c\in \mathcal{C}}{\frac{m^i_c}{\textrm{SNR}^{\textrm{th}}_{c}}}, \hspace{10em}\forall i.
\end{align}
\end{subequations}

Constraints (\ref{eqn: ASE}) calculate PLI of ASE noise on the lightpath. Also, it is applied on the SCI calculation  in constraints (\ref{eqn: SCI}). Whether request $i$ uses link $uv$ and transmission mode $c$ is assured by the constraints (\ref{eq: flowlinktm}).

\begin{figure}[!htbp]
    \centering
\includegraphics[width=6.5cm]{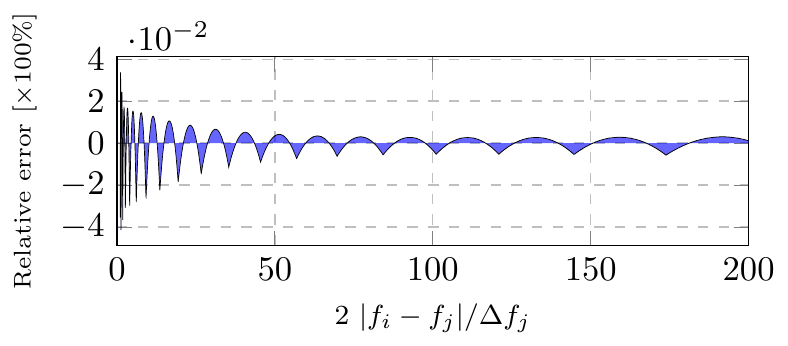}
    \caption{Illustration for the piece-wise linear fitting performance, $(\hat{h}_c^{ij} - h_c^{ij})/h_c^{ij}$ versus $2|f_i-f_j|/\Delta f_j$, $Q$=20.}
    \label{fig: error_pwl}
\end{figure}

Since XCI is caused by two lightpaths, it will increase along their sharing links, and non-decrease along the other links. Constraints (\ref{eqn: XCIij}) and (\ref{eqn: XCIi}) implement the XCI calculation, respectively. But the nonlinear expression between XCI and $f_i$ in Eq.~(\ref{eq: GiXCI}) makes a nonlinear calculation term $h_c^{ij}$. To address this issue, we replace the nonlinear term $h_c^{ij}$ by the following linear approximation $\hat{h}_c^{ij}$,
\begin{align}
&h^{ij}_c \left(2\frac{|f_i-f_j|}{\Delta f_j} \right) \geq \ln{\left(\frac{2|f_i-f_j|/\Delta f_j + 1}{2|f_i-f_j|/\Delta f_j - 1}\right)}   \\
\Leftarrow & \hat{h}_c^{ij}(x) \geq \max \left( o_1^1 x + o_1^0, \cdots, o_k^1 x + o_k^0, \cdots, o_Q^1 x + o_Q^0\right) \nonumber \\
\Leftarrow & \hat{h}_c^{ij} \geq o_k^1(2 \textrm{SE}(c) f_{ij}/ r_j )+ o_k^0 \hspace{3em} \forall i\neq j,\forall c,1\leq k\leq Q \nonumber
\end{align}
where $o_k^1$ and $o_k^0$ are the coefficients solved by piece-wise linear fitting. We use the least-square algorithm in \cite{MaBo09} to fit the convex function  $\ln \left(\frac{x+1}{x-1}\right)$ in the domain $x\in [x_1, x_2]$, where we set $x_1$=1.001, $x_2$=200, and $Q$=20. The fitting error ($\hat{h}_c^{ij} - h_c^{ij})/h_c^{ij}$ can be minimized by increasing the number of segments $Q$. As shown in Fig. \ref{fig: error_pwl}, the maximum fitting error is less than $5\%$.

Constraints (\ref{eqn: conPLIAD}) calculate the node crosstalk along the lightpath, which is implemented by emphasizing the incoming degree $q_v^j$ of crosstalk signal $j$ and the outgoing degree $p_v^i$ of primary signal $i$. Constraints (\ref{eqn: conPLI5}) calculate the total PLIs of all traversed links and nodes. Constraints (\ref{eqn: conPLI7}) represent the QoT formulation. The MILP is NP hard and time-consuming\cite{ChMV10}. Considering the scalability limitations of MILP, we also design a heuristic algorithm to solve the revenue maximization problem with lower complexity.

\section{Heuristic Algorithms}\label{sec: heu}

In this section, we design a decomposition algorithm (DEC-ALG) to solve (a) the problem of routing, MF, and FEC assignment, and (b) the problem of spectrum assignment, separately. In addition, we also present a heuristic as benchmark, which is adapted from existing literature.

\subsection{DEC-ALG}

DEC-ALG algorithm consists of two phases: (a) Routing and transmission mode assignment (RTMA); (b) Spectrum assignment (SA). The related parameters and variables are explained in Table \ref{tab: ParaHeu}.

\begin{table}[!htbp]
\caption{Parameters \& Variables in algorithm DEC-ALG}
\label{tab: ParaHeu}
\scalebox{0.95}{
\begin{tabular}{p{1.6cm}p{6.8cm}}
\toprule
\multicolumn{2}{c}{\textbf{Network sets \& Parameters}}\\
\midrule
$\phi \in [0,1]$  & A ratio on SNR threshold considering ASE+SCI. $1-\phi$ represents the part considering XCI+AD.\\
$P_{i}$ & Route and transmission mode pairs set $(\xi, c)$ of request $i$.\\
$P_{ie}$  & Route and transmission mode pairs set $(\xi,c)$ of request $i$ that traverses link $e$.\\
$V_{i\xi}$  & Node set on the $\xi$-th route of request $i$.\\
$\epsilon_2$ & Factors adjusting the weight of rest SNR margin.\\
$b_i,e_i$ & Spectrum beginning and end of request $i$.\\
$\phi_{i\xi c}\in \mathbb{R} $ & SNR margin of request $\xi$ using transmission mode $c$ and $\xi$-th route after RTMA.\\
$\Delta f_{ic} $ & Bandwidth of request $i$ using transmission mode $c$.\\
$N_{RTMA}$ &  Number of RTMA$_\text{opt}$ solutions.\\
$N_\textrm{round}$ & Number of attempts in the spectrum assignment.\\
\midrule
  \multicolumn{2}{c}{\textbf{Variables}}\\
  \midrule
$B_i\in\{0,1\}$ & Equals 1 if request $i$ is accepted, 0 otherwise.\\
$g_{i\xi c}\in\{0,1\}$ & Equals 1 if request $i$ uses the $\xi$-th route and transmission mode $c$, 0 otherwise.\\
$\phi_{\textrm{avg}} $ & Average SNR margin of all requests.\\
\bottomrule
\end{tabular}
}
\end{table}

\subsubsection{Pre-calculation process}
We pre-calculate route and transmission mode pairs for each request. Candidate routes are obtained by the $K$ shortest path algorithm\cite{Yen71}, while transmission modes are from $\mathcal{C}$. 

For any request $i$, the tuple ($\xi_i,c_i$) denotes a route and transmission mode pair. The ASE and SCI are the known impairments for a given ($\xi_i,c_i$), which are obtained by Eqs. (\ref{eq: GiASE}) and (\ref{eq: GiSCI}). However, the node crosstalk and XCI can not be determined because they are related with the exact spectrum channel of all lightpaths. Thus, some route and transmission mode pairs may experience strong PLIs, thus the QoT cannot be satisfied in the subsequent process. Therefore, we choose the route and transmission mode pair ($\xi,c$) based on the residual SNR margin $\phi_{i\xi c}$ as follows,
\begin{equation} \label{eq: phiirc}
\phi_{i\xi c} = \frac{\phi}{ \textrm{SNR}_c^{\textrm{th}}} - t_i^{\textrm{ASE}} - t_i^{\textrm{SCI}} - \epsilon_{\textrm{X}} \sum_{v\in V_{i\xi}} \frac{N(v)+1}{2}
\end{equation}
where $\phi$ is an estimated ratio considering the ASE and SCI.

\subsubsection{RTMA}

In RTMA, each request can choose one route and transmission mode pair ($\xi,c$). For all requests, one assignment of the route and transmission mode pair, named as RTMA$_\textrm{opt}$, is  solved by the following RTMA model,
\begin{equation*}
\max \quad  \sum_{i\in D}  \eta_i B_i  + \epsilon_2 \phi_{\textrm{avg}}  \quad \boldsymbol{\mathrm{(RTMA)}}
\end{equation*}
\begin{subequations}
\label{model: RTMA}
\begin{align}
\label{eq: g_irc}
\textbf{\textit{s.t.}} & \sum_{(\xi,c)\in P_i} g_{i\xi c} = B_i &&\forall i\\
\label{eq: FIBER}
& \sum_{i \in D} \sum_{(\xi,c)\in P_{ie}} \Delta f_{ic} g_{i\xi c} \leq F &&\forall e\\
\label{eqn: phiu}
& 0 \leq \sum_{(\xi,c) \in P_i} g_{i\xi c} \phi_{i\xi c}, && \forall i\\
\label{eq: phia}
& \phi_{\textrm{avg}} = \frac{1}{|D|} \sum_{i\in D} \sum_{(\xi,c) \in P_i} g_{i\xi c} \phi_{i\xi c} &&
\end{align}
\end{subequations}

The main objective of RTMA is to maximize the accepted revenue and the second one is to maximize average SNR margin. The multi-objective function  can be adjusted by weighting factor $\epsilon_2$. Constraints (\ref{eq: g_irc}) make sure that a route and transmission mode pair ($\xi,c$) is chosen for request $i$ if $B_i$=1. Constraints (\ref{eq: FIBER}) restrict the spectrum usage of each link. Bandwidth requirement of $i$ with the transmission mode $c$ has been pre-calculated and denoted by $\Delta f_{ic}$.
Constraints (\ref{eqn: phiu}) make sure that the minimum SNR margin is non-negative. Constraint (\ref{eq: phia}) defines the average SNR margin of all requests.

In RTMA, only one RTMA$_{\textrm{opt}}$ solution is obtained. However, this solution may not bring the maximum revenue after spectrum assignment. To this end, we intend to generate $N_{RTMA}$ solutions. The $n$-th solution is generated by adding constraints (\ref{eqn: loop}) to RTMA, which is used for excluding the previous RTMA$_{\textrm{opt}}$. In constraints (\ref{eqn: loop}), both $B_i^{n-1}$ and $g_{i\xi c}^{n-1}$ are the results from $(n-1)$-th solution.


\begin{equation}\label{eqn: loop}
\begin{split}
&\sum_{\substack{i\in D \\ B_i^{n-1}=1}} \sum_{\substack{(\xi,c)\in P_i \\ g_{i\xi c}^{n-1}=1 } } g_{i\xi c} + \frac{1}{K*|\mathcal{C}|} \sum_{\substack{i\in D \\
B_i^{n-1}=0}} 
\sum_{\substack{(\xi, c)\in P_i \\ g_{i\xi c}^{n-1}=0 } } (1-g_{i\xi c} ) \\ 
& \leq |D|- \frac{1}{K*|\mathcal{C}|}, n \in \{1, 2, 3, \cdots, N_{RTMA}\} \end{split}
\end{equation}

\textbf{Explanation of excluding constraints (\ref{eqn: loop})} : In the $n$-th loop, we suppose that $\forall i,(\xi,c)$, the variable $g_{i\xi c}=g_{i\xi c}^{n-1}$, then the value of the left side in constraints (\ref{eqn: loop}) becomes $|D|$, which is larger than the right. Therefore, we can say that constraints (\ref{eqn: loop}) hold if $\exists i,(\xi,c), g_{i\xi c}\neq g_{i\xi c}^{n-1}$. 
Let us focus on the request $i$ that satisfies $g_{i\xi c}\neq g_{i\xi c}^{n-1}$. Since constraints (\ref{eq: g_irc}) require  $\sum_{(\xi,c)} g_{i\xi c}^{n-1} = B_i^{n-1}\leq 1$, we discuss the case $B_i^{n-1}=1$ and  $B_i^{n-1}=0$, respectively. 
\begin{itemize}
    \item If $B_i^{n-1} = \sum_{(\xi,c)} g_{i\xi c}^{n-1} = 1$ holds,
    \end{itemize}
\begin{align}
\left. 
\begin{aligned}
\label{eq: n0}
B_i^{n-1}=\sum_{(\xi,c)} g_{i\xi c}^{n-1}=1\\
\exists (\xi,c), g_{i\xi c}\neq g_{i\xi c}^{n-1}\\
\sum_{(\xi,c)} g_{i\xi c} \leq 1
\end{aligned}\right\}\Rightarrow \sum_{\substack{(\xi,c)\in P_i \\ g_{i\xi c}^{n-1}=1 } } g_{i\xi c} = 0
\end{align}
\begin{itemize}
    \item Otherwise $B_i^{n-1} = \sum_{(\xi,c)} g_{i\xi c}^{n-1} = 0$ holds, then we can get the result     $\frac{1}{K*|\mathcal{C}|} \sum_{\substack{(\xi,c)\in P_i \\ g_{i\xi c}^{n-1}=0}} (1 - g_{i\xi c})$=$ \frac{K*|\mathcal{C}|-1}{K*|\mathcal{C}|}$<1 with the following proof,\noindent
    \end{itemize}
\begin{align}
\left. 
\begin{aligned}
\label{eq: n1}
B_i^{n-1}=\sum_{(\xi,c)} g_{i\xi c}^{n-1}=0\\
\exists (\xi,c), g_{i\xi c}\neq g_{i\xi c}^{n-1}\\
\sum_{(\xi,c)} g_{i\xi c}\leq 1
\end{aligned}\right\}& \Rightarrow
\left\{
\begin{aligned}
&\sum_{\substack{(\xi,c)\in P_i \\ g_{i\xi c}^{n-1}=0 } } g_{i\xi c}=1 \\
&\sum_{\substack{(\xi, c)\in P_i \\ g_{i\xi c}^{n-1}=0 } } (1-g_{i\xi c}) = K*|\mathcal{C}|-1
\end{aligned}\right.
\end{align}

In addition, for the request $i$ that satisfies $\forall (\xi,c), g_{i\xi c}= g_{i\xi c}^{n-1}$, the first item of left sides still equals 1. The number is denoted by $n_1$, $n_1\leq |D|-1$. Thus, we can get $0\cdot n_{(\ref{eq: n0})} + 1 \cdot n_1 + (1-1/K/|\mathcal{C}|)  \cdot n_{(\ref{eq: n1})} \leq |D|-1/K/|\mathcal{C}|$, where $n_{(\ref{eq: n0})}$ and $n_{(\ref{eq: n1})}$ are the number of requests satisfying (\ref{eq: n0}) and (\ref{eq: n1}), respectively, and  $n_{(\ref{eq: n0})}+n_1+n_{(\ref{eq: n1})}=|D|$. Therefore, the left side of constraints (\ref{eqn: loop})  must be no bigger than $|D|-\frac{1}{K\cdot |\mathcal{C}|}$. 

The pseudo code in Algorithm \ref{alg: RTMA} illustrates the procedure of generating $N_{RTMA}$ solutions by RTMA. In line~\ref{alg: RTMA_1}, the model is initialized with the constraints (\ref{model: RTMA}) and the input parameters $\mathrm{G}(V,E)$, $D$, $N_{RTMA}$, and $\mathcal{C}$. In line~\ref{alg: RTMA_2}, RTMA$_{\textrm{opt}}$  solution with $g_{i\xi c}=0$ is initialized. Then, from lines~\ref{alg: RTMA_For} to \ref{alg: RTMA_For_End}, the RTMA model is repeated to get $N_{RTMA}$ solutions.

\begin{algorithm}[!htbp]
\SetKwInOut{Input}{Input}
\SetKwInOut{Output}{Output}
\SetKw{KwAnd}{and}
\SetKw{KwSuch}{s.t.}
\Input{$\mathrm{G(V,E)}, D, N_{RTMA},  \mathcal{C}$}
\Output{$\boldsymbol{g^{N_{RTMA}}}$}
Create the RTMA model with the constraints in (\ref{model: RTMA}) and the input parameters $\text{G}$(V,E), $D$,  $N$, and $\mathcal{C}$\label{alg: RTMA_1}\;
$g^0_{i\xi c} \gets 0$,  $\forall i, (\xi, c)$ \tcp{$g_{i\xi c}^0\in \boldsymbol{g^0}$}\label{alg: RTMA_2}
\For{$n\in \{1,2,...,N_{RTMA}\}$}
{\label{alg: RTMA_For}
Update the RTMA model with excluding constraints (\ref{eqn: loop}) and previous solution $\boldsymbol{g^{n-1}}$\;
Get the RTMA$_{\textrm{opt}}$ solution $\boldsymbol{g^{n}}$ by solving RTMA model\label{alg: RTMA_For_End}\;
}
\caption{RTMA: generating $N_{RTMA}$ solutions}
\label{alg: RTMA}
\end{algorithm}



\subsubsection{SA}
Once the RTMA problem is solved, from the solution RTMA$_{\textrm{opt}}$, we can obtain the pair index of route and transmission mode used for each request $i$, \textit{i.e.}, $(\bar{\xi_i}, \bar{c_i})= \{(\xi, c)| g_{i\xi c}= 1,i\in D\}$. 

When assigning the spectrum interval on the determined route for each request, SA needs to take into account both spectrum continuity and spectrum contiguity constraints. To reduce the impact of XCI, we also set a guard band $\Delta$=12.5 GHz. If the request is accepted, a specific lightpath with its spectrum interval will be allocated. Otherwise, it will move the spectrum interval when necessary until it is out of the fiber spectrum. An new incoming request can be blocked if it affects the QoT of other requests. To ensure the blocked requests can be accepted again, we repeat the assignment process $N_{\text{round}}$ times. For the request in each round, the SNR threshold  is designed to decrease gradually, and equals to the SNR threshold of transmission mode $c$ in the final round.

The SA procedure is illustrated in Algorithm \ref{alg: SA}. In line~\ref{alg: SA_arr}, we sort the requests in $D$ by function $\textrm{ARRANGE}$, which will be explained later. Then, in lines~\ref{alg: SA_r} and \ref{alg: SA_fi}, the required spectrum bandwidth and route can be both found from the RTMA result RTMA$_\text{opt}$. In line~\ref{alg: SA_merge}, we merge the available spectrum of the $\xi$-th route and assign it to \textit{MERGED}$_{\textrm{space}}$,  where both spectrum continuity and spectrum contiguity constraint are satisfied. Then, from lines~\ref{alg: SA_scan_start} to \ref{alg: SA_scan_end}, the algorithm SA tries to search a spectrum interval [$b_i$, $b_i$ + $\Delta f_i$] from \textit{MERGED}$_{\textrm{space}}$ that satisfies the QoT.

In the ARRANGE function of Algorithm \ref{alg: SA}, we sort the requests in $D$ by four different assignment polices, \textit{random order} $\Delta f_i$ (SA), descending order of \textit{bandwidth} $\Delta f_i$ (SA-B), \textit{revenue} $\eta_i$ (SA-R), and \textit{revenue to bandwidth ratio} $\eta_i/\Delta f_i$ (SA-RA), respectively. The performances of these four arrangement policies are compared in simulations.

\begin{algorithm}[!htbp]
\SetKwInOut{Input}{Input}
\SetKwInOut{Output}{Output}
\SetKw{KwAnd}{and}
\SetKw{KwSuch}{s.t.}
\SetKwProg{try}{try}{:}{}
\SetKwProg{catch}{catch}{:}{end}
\Input{$\boldsymbol{g}$}
\Output{$\boldsymbol{alloc}$}
$D_{\textrm{arr}} \gets \textbf{ARRANGE}(D)$\;\label{alg: SA_arr}

$\boldsymbol{alloc} \gets \boldsymbol{0}$ \tcp{$alloc_i \in \boldsymbol{alloc}$;}

\For{ n$_{round}\in$ $[0,1,...,N_{ROUND}]$}
{

\For{$i \in D_{\textrm{arr}} $ \& alloc$_i$==0}
{
$(\bar{\xi_i}, \bar{c_i})  \gets \{(\xi,c) |g_{i\xi c}=1\} $\label{alg: SA_r}\tcp{$g_{i\xi c} \in \boldsymbol{g}$;}
$\Delta f_i \gets r_i/\textrm{SE}(\bar{c_i})$\label{alg: SA_fi} \;
MERGED$_{\textrm{space}}$ $\gets $ available spectrum space on $\bar{\xi_i}$-th route\label{alg: SA_merge}\;

\For{the spectrum beginning $b_i$: 0$\rightarrow$F, with step $\Delta$=12.5 GHz \& alloc$_i$==0\label{alg: SA_scan_start}}
{
 \If{$[b_i, b_i + \Delta f_i]$ $\in$ $\mathrm{MERGED}_{\mathrm{space}}$ }
 { 
 \try{ 
 }{
 Assign $[b_i, b_i + \Delta f_i ]$ on $\bar{\xi_i}$-th route\;
Check QoT of assigned requests by Eq. (\ref{eq: OSNR newform})\;
Check QoT of current request $i$ by Eq. (\ref{eq: OSNR newform})\;
    $alloc_i$ $\gets 1$\;
 }
 \catch{Check failed}{
    $alloc_i$ $\gets 0$\;\label{alg: SA_scan_end}
 }
 }
 }
}

}

\caption{SA : Spectrum Assignment}\label{alg: SA}
\end{algorithm}

\subsubsection{DEC-ALG} It is worth mentioning that the size of RTMA$_\textrm{opt}$ solution space can reach $\left( K\cdot |\mathcal{C}|+1\right)^{|D|}$. With small $N_{RTMA}$, \textit{i.e.} $N_{RTMA}\ll\left(K\cdot |\mathcal{C}|+1\right)^{|D|}$, the potential RTMA$_\textrm{opt}$ solution that provides the maximum revenue may not be included. 
In addition, the solution space of RTMA is determined by the number of routes $K$, the size of candidate transmission mode $\mathcal{C}$ and demand $D$. It may happen that we get the identical result even with different configurations $\mathcal{C}'$ or different demand matrix $D'$.

To solve the problem caused by different transmission mode configurations $\mathcal{C}$, we use a perturbation strategy to extend RTMA$_\textrm{opt}$ solutions by its subset, $\mathcal{C}^{sub}_l$, $1 \leq l\leq |\mathcal{C}|$. Each subset takes $l$ elements from $\mathcal{C}$. For the $l$-th subset $\mathcal{C}_l^{sub}$, the $l$-th transmission mode of $\mathcal{C}$ is added compared to the previous $l-1$ subsets. Constraints (\ref{eqn: lock}) are used to generate RTMA$_\textrm{opt}$ for the subset $\mathcal{C}_l^{sub}$ that forces the use of transmission mode $c^l$ but excludes the use of the other transmission modes $c^r$.
\begin{subequations}\label{eqn: lock}
\begin{align}
&\sum_{ i\in D, 1\leq \xi\leq K} g_{i\xi c^l} \geq 1 \\
&\sum_{i\in D, 1\leq \xi\leq K} g_{i\xi c^r} =0, \forall c^r \in \mathcal{C}	\textbackslash \mathcal{C}^{sub}_l
\end{align}
\end{subequations}

We give the pseudo code of DEC-ALG as illustrated in Algorithm \ref{alg: RMFSA}. 
In line~\ref{alg: init_subset}, we initialize the candidate transmission sets $\mathcal{C}^{sub}_l$. From lines~\ref{alg: RMTA_inDECALG} to \ref{RF: rev}, we conduct the algorithm RTMA and SA to obtain the optimal result. Finally, the maximum revenue $\text{MAXOBJ}$ is saved.


%


\begin{algorithm}[!htbp]
\SetKwInOut{Input}{Input}
\SetKwInOut{Output}{Output}
\SetKw{KwAnd}{and}
\SetKw{KwSuch}{s.t.}
\Input{$\mathrm{G}(V,E), D, N_{RTMA}, \mathcal{C}$}
\Output{MAXOBJ}
MAXOBJ $\gets$ 0\;
\For{$c^l \in \mathcal{C}$ }{
\label{alg: init_subset}
Initialize the current transmission mode set $\mathcal{C}^{sub}_l$\;
\label{alg: RMTA_inDECALG}
Use RTMA model to generate RTMA$_{\textrm{opt}}$(G, $D$, $N_{RTMA}$, $\mathcal{C}$) solutions, with constraints (\ref{eqn: lock}) emphasising the lock transmission mode $c^l$\label{alg: gnirc}\;

\For{$n \in \{0,1,...,N_{RTMA}\}$}
{
  $\boldsymbol{alloc} \gets \textrm{SA}(\boldsymbol{g}^n)$ \label{RF: alloc} \tcp{\textrm{alloc}$_i\in \boldsymbol{alloc}$}
  Obj $\gets$ $\sum_i$
  alloc$_i$*$\eta_i$ \label{RF: rev}\;
  MAXOBJ$\gets$$\max$(Obj, MAXOBJ)\;
}


}
\caption{DEC-ALG}\label{alg: RMFSA}
\end{algorithm}
\subsection{Benchmark algorithm}

To efficiently utilize the spectrum resource of fiber, a large number of algorithms on traffic provisioning have been proposed\cite{YADW17,ZhWA15}. To make a fair comparison, we take the algorithm in \cite{YADW17} that also adopts the continuous spectrum allocation. The benchmark algorithm, called as REF-A in this paper, is implemented by using the same objective function of RTMA in Eq. (\ref{model: RTMA}) and restricting the fiber spectrum resources. It should be also noted that, the spectrum assignment of REF-A is implemented by passing the solution of RTMA to RMAX.



\section{Illustrative Numerical Results}\label{sec: simulation}

In this section, we present the numerical experiment results. First, we compare the efficiency of our proposed heuristic and the MILP model. Then, we investigate revenues in scenarios with different PSDs and different transmission modes. Finally, we consider the experiments for severe resource crunch, which is simulated by increasing bit-rate and number of requests.

The MILP, heuristic algorithm DEC-ALG, and REF-A run on an Intel Core PC with 4.0~GHz CPU and 16~GB RAM. Specifically, we solved the MILP model by CPLEX 12.6 and implemented the two heuristic algorithms using an ad-hoc code developed in C++. Maximum computing time for the MILP was fixed to one hour. All illustration results have been averaged over 10 independent simulation runs to guarantee statistical accuracy.


The 6-node network in Fig. \ref{fig: instance}, NSFNET (14 nodes, 44 links) \cite{JZZX16}, and US Backbone network (28 nodes, 90 links)\cite{JZZX16} are used as case study topologies (note that, since the path length of NSF network and US Backbone network cannot support high-order MF, we divide the  length of link by 6 in the simulations). The spectrum resource of each fiber $F$ is assumed with 1,000~GHz to increase the simulation speed for large networks. The fiber parameters $\alpha$, $\beta_2$, and $\gamma$ are from Table \ref{tab: pli_notations}. The algorithm parameter $\epsilon_1$=0.01 and  $\epsilon_{2}$=0.001 are adjusted to be small to emphasize the revenue rather than the other parameters for simulation. The parameters $N_{RTMA}$=40, $N_\textrm{round}$=2, and $K$=4 are adjusted to guarantee stable good simulation results in a reasonable time.
The bit-rates $r_{i}$ are randomly chosen from the set $\{$250, 500, ... , 250+$n$*250, ... 250+2$n$*250$\}$~Gbps. For the lowest bit-rate, the channel can be guaranteed with  bandwidth over than 28~GHz with PM-16QAM\cite{YADW17}, which is acceptable for the GN model in \cite{JoAg14}. The large bit-rate request is assumed by super-channel with large baud rates. The initial launch power PSD for all lightpaths is simplified  with -16 dBm/GHz by using the LOGON strategy \cite{PBCC13} for one span with the heaviest spectral loads. The definition of revenue and other used notations for simulation are given as follows,

\begin{enumerate}
\item \textit{Revenue}: $\eta_i =  u_i$, where $u_i$ is the service type parameter. In this paper, we consider the service type parameter $u_i$ follows the Zipf distribution Zipf(1,5)\cite{ZLSB15}. 
The revenue of a network is sum of all accepted lightpaths' revenue.
\item \textit{Adaptive MFs} : $\mathcal{C}=(\hat{\mathcal{M}}_m,\mathcal{F}_f)$. Notation $\mathcal{F}_f$ represents one $f$-th level FEC, and  $\hat{\mathcal{M}}_m$ represents all the MFs not beyond the $m$-th order.
\item \textit{Multiple FECs}: $\mathcal{C}=(\mathcal{M}_m,\hat{\mathcal{F}}_f)$. Notation  $\mathcal{M}_m$ represents the $m$-th order MF, and $\hat{\mathcal{F}}_f$ represents the FEC OHs not beyond the $f$-th one.

\end{enumerate}

\subsection{Validation using MILP}

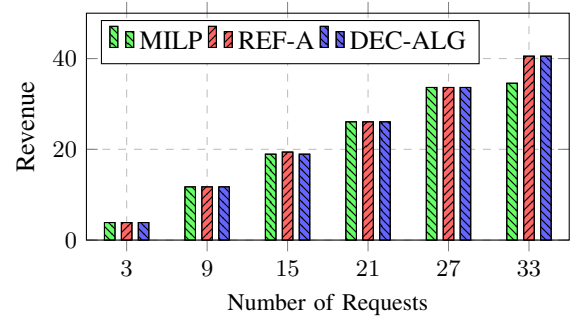
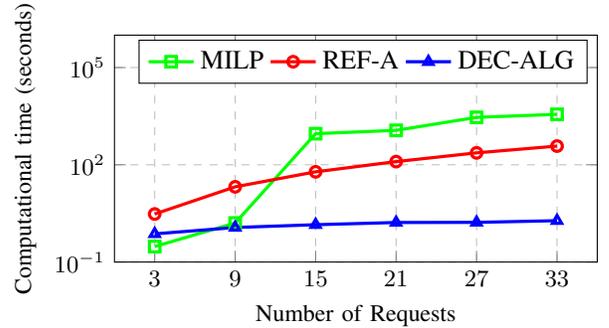
\begin{figure}[!htbp] 
\centering

\subfigure[Revenue]{
\begin{tikzpicture}
\begin{axis}[
       label style={font=\small},
       tick label style={font=\small},
	width=8cm,
	height=4.6cm,
	ymin=0, ymax=50,
	xmin=0, xmax=36,
	xtick = {3,  9,  15, 21, 27, 33},
	ylabel near ticks,
	ybar=6*\pgflinewidth,
	bar width=4pt,
	xlabel={Number of Requests},
	ylabel={Revenue},
    legend pos= north west,
    legend columns=-1,
    xmajorgrids,
    ymajorgrids,
    grid style = dashed,
]
\pgfplotstableread{TON/Results_RevTimevsNoR.txt}\loadedtable
\addplot[
black!60!black,
fill=white!40!green,
postaction={
    pattern=north west lines
}]table [x index=0, y index=1] {\loadedtable};
\addplot[black!60!black,
fill=white!40!red,
postaction={
    pattern=north east lines
}]table [x index=0, y index=2] {\loadedtable};
\addplot[black!60!black,
fill=white!40!blue,
postaction={
    pattern=north west lines
}]table [x index=0, y index=3] {\loadedtable};
\legend{MILP,REF-A,DEC-ALG};
\end{axis}
\end{tikzpicture}
}
\subfigure[Computational time]{
\begin{tikzpicture}
\begin{axis}[
       label style={font=\small},
       tick label style={font=\small},
	width=8cm,
	height=4.6cm,
    ymin=0.1, ymax=1000000,
	xtick = {3,  9,  15, 21, 27, 33},
    ymode=log,
	xlabel={Number of Requests},
    ylabel={Computational time (seconds)},
    ylabel style={yshift=0.0cm},
    legend columns=-1,
    xmajorgrids,
    ymajorgrids,
    grid style = dashed,
]
\pgfplotstableread{TON/Results_RevTimevsNoR.txt}\loadedtable
\addplot[mark=square,white!00!green,very thick] table [x index=0, y index=4] {\loadedtable};
\addplot[mark=o,white!00!red,very thick] table [x index=0, y index=5] {\loadedtable};
\addplot[mark=triangle,white!00!blue,very thick] table [x index=0, y index=6] {\loadedtable};
\legend{MILP,REF-A,DEC-ALG};
\end{axis}
\end{tikzpicture}
}
\caption{Comparison of revenue and computational time in 6-node network. }\label{fig: 6node_heu_comp}
\end{figure}

We validate the MILP on the 6-node network. The bit-rate per request is fixed at 1,000~Gbps. Figure \ref{fig: 6node_heu_comp} illustrates the revenue and computational time of three algorithms as the number of request increases. 

In Fig.~\ref{fig: 6node_heu_comp}, we can observe that the computational time of MILP reaches the preset maximum computing time one hour, when the number of requests increases to 33. It means that MILP is intractable even in the case with either small networks or small number of requests. But the heuristic algorithm REF-A and DEC-ALG can solve it in a few minutes and a few seconds, respectively. Besides, we observe that both DEC-ALG and REF-A are able to obtain an approximate optimal value of MILP. Therefore, the proposed algorithm DEC-ALG is not only time-efficient but also near-optimal.

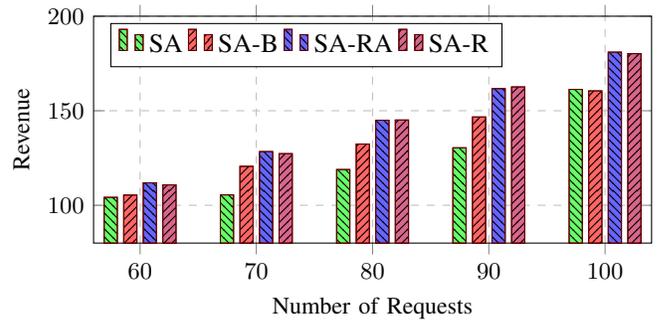
\begin{figure}[!htbp]
\flushleft
\begin{tikzpicture}
\begin{axis}[
width = 9 cm,
height = 4.6cm,
bar width=5pt,
ybar=6*\pgflinewidth,
ymin = 80, ymax = 200,
xtick = {60,70,80,90,100},
ylabel = {Revenue},
ylabel near ticks,
xlabel = {Number of Requests},
xmajorgrids,
ymajorgrids,
grid style = dashed,
legend pos = north west,
legend columns = 5,
       label style={font=\small},
       tick label style={font=\small},
]
\pgfplotstableread{TON/ARRANGE_Metric.txt}\loadedtable
\addplot[black!60!red, fill = white!40!green, postaction = {pattern = north west lines}
] table [x index = 0, y index = 1] {\loadedtable};

\addplot[black!60!red, fill=white!40!red, postaction = {pattern = north east lines}
] table [x index =0, y index = 2] {\loadedtable};

\addplot[black!60!red, fill=white!40!blue, postaction = {pattern = north west lines}
] table [x index =0, y index = 3] {\loadedtable};

\addplot[black!60!red, fill=white!40!purple, postaction = {pattern = north east lines}
] table [x index =0, y index = 4] {\loadedtable};

\legend{SA, SA-B, SA-RA, SA-R};
\end{axis}
\end{tikzpicture}

\caption{Revenue comparison with four different sorting policies of ARRANGE function.}
\label{fig: 3heuristic}
\end{figure}

In the SA of DEC-ALG algorithm, we have mentioned four different sorting policies in the ARRANGE function. In order to find the best sorting policy, we compared their results in Fig. \ref{fig: 3heuristic}. The simulation is carried out in NSF network. As we see in Fig. \ref{fig: 3heuristic}, the revenue with different sorting policies increases with the number of requests. It can be also seen that SA-RA, which sorts the requests by the descending order of revenue/bandwidth ratio, gets the largest revenue. Therefore, we confirm to use SA-RA for heuristic algorithm DEC-ALG.


\subsection{Impacts of PSD, MF and FEC}

\begin{figure}[!htbp]
\centering
\scalebox{0.9}{
\begin{tikzpicture}

\begin{axis}
[
width=8cm,
height=4.8cm,
ymin=-1,  ymax= 30,
xmin=-36, xmax=-7,	
ylabel near ticks,
xlabel={PSD $G_i$ [dBm/GHz]},
ylabel={SNR [dB]},
legend pos= south east,
legend columns=4,
xmajorgrids,
ymajorgrids,
grid style =dashed,
grid = both,
]
\pgfplotstableread{TON/SNRvsPSD.txt}\loadtable
\addplot[mesh, very thick] table [x = PSD, y  = SNR]{\loadtable};

\node[text width = 3cm] at (4.5cm, 3cm) {1-low SNR};
\node[text width = 3cm] at (6.7cm, 3cm) {2-high SNR};
\node[text width = 3cm] at (8.7cm, 3cm) {3-low SNR};
\end{axis}

\begin{axis}
[
    width=8cm,
	height=4.8cm,
	axis y line* = right,
axis x line = none, 
ymin = 1e-7, ymax = 4e-5, 
xmin = -36, xmax = -7,
ylabel = {Interference or noise [W]},
legend pos = south west,
      label style={font=\small},
      tick label style={font=\small},
]
\pgfplotstableread{TON/SNRvsPSD.txt}\loadtable
\addplot[blue, thick,dashed] table [x = PSD, y = ASE]{\loadtable};
\addplot[red, thick, dotted] table [x = PSD, y = SCI]{\loadtable};
\addplot[cyan, thick,densely dashed] table [x = PSD, y  = XCI]{\loadtable};
\legend{ASE, SCI, XCI}
\end{axis}
\end{tikzpicture}
}
\caption{SNR vs. PSD $G_i$.}
\label{fig: SNR_PSD}
\end{figure}
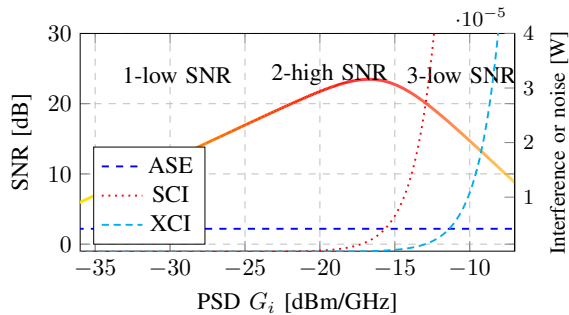

As we have seen in the example of Fig. \ref{fig: instance},  revenues can be influenced by SNR requirements of different transmission mode  configurations. In Eqs. (\ref{eq: OSNR newform}), (\ref{eq: GiASE}) and (\ref{eq: GiSCI}), when PSD $G_i$ increases, the noise to signal ratio of ASE, $t^{\textrm{ASE}}_i$ will decrease inversely, while the $t^{\textrm{SCI}}_i$ and $t^{\textrm{XCI}}_i$ will increase quadratically. 
The PSD of ASE noise, SCI, and XCI, as well as the SNR for a 250~Gbps request with PM-QPSK$_{7\%}$ in the middle of a fully occupied fiber span are illustrated in Fig.~\ref{fig: SNR_PSD}.
According to the SNR and PSD, we briefly distinguish three different scenarios, namely scenario 1:~\textit{low SNR with low PSD}, scenario 2:~\textit{high SNR with median PSD}, and scenario 3:~\textit{low SNR with high PSD}. By adjusting PSDs, we can investigate the revenue impact of MF and FEC in different SNR scenarios.

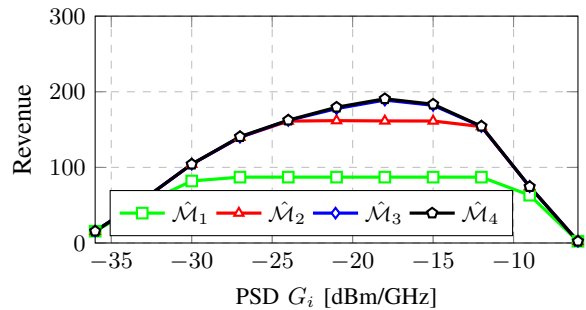
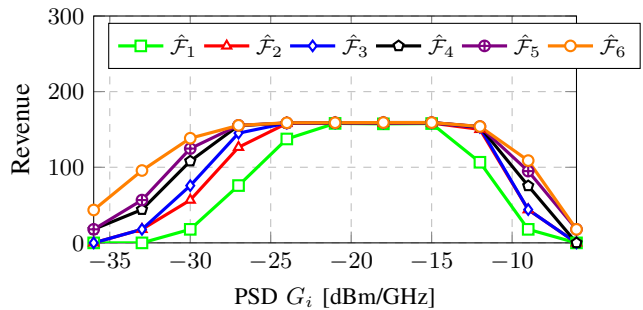
\begin{figure}[!htbp]
\centering
\subfigure[Different adaptive MF configurations]{
\label{fig: NSF_MF_power}
\begin{tikzpicture}
\begin{axis}[
       label style={font=\small},
       tick label style={font=\small},
width = 8cm,
height = 4.6cm,
ymin = 0, ymax = 3e2,
xmin = -36, xmax = -6,
ylabel near ticks,
xlabel ={PSD $G_i$ [dBm/GHz]},
ylabel ={Revenue},
legend pos =south west,
legend columns = 5,
legend style={font =\footnotesize},
xmajorgrids,
ymajorgrids,
grid style = dashed]
\pgfplotstableread{TON/NSF_MF_power.txt}\loadtable
\addplot[mark=square*, mark options = {fill =white, thick}, white!0!green, very thick] table [x index=0, y index = 1] {\loadtable};
\addplot[mark=triangle*, mark options = {fill = white, thick}, white!0!red,  very thick] table [x index=0, y index = 2]{\loadtable};
\addplot[mark=diamond*, mark options = {fill = white, thick}, white!0!blue,  very thick] table [x index=0, y index = 3]{\loadtable};
\addplot[mark=pentagon*, mark options = {fill = white, thick}, white!0!black,  very thick] table [x index=0, y index = 4]{\loadtable};
\legend{$\mathcal{\hat{M}}_1$,$\mathcal{\hat{M}}_2$, $\mathcal{\hat{M}}_3$,$\mathcal{\hat{M}}_4$};
\end{axis}
\end{tikzpicture}
}
\subfigure[Different  multiple FEC configurations]{
\label{fig: NSF_FEC_power}
\begin{tikzpicture}
\begin{axis}[
label style={font=\small},
tick label style={font=\small},
width = 8cm,
height = 4.6cm,
ymin = 0, ymax = 3e2,
xmin =-36, xmax = -6,
ylabel near ticks,
xlabel={PSD $G_i$ [dBm/GHz]},
ylabel={Revenue},
legend pos = north west,
legend style ={font=\footnotesize},
legend columns =7,
xmajorgrids,
ymajorgrids,
grid style =dashed
]
\pgfplotstableread{TON/NSF_FEC_power.txt}\loadtable

\addplot[mark = square*, mark options = {fill =white, thick}, white!00!green, very thick] table [x index = 0, y index = 1] {\loadtable};
\addplot[mark = triangle*,mark options = {fill= white, thick}, white!00!red, very thick] table [x index = 0, y index = 2] {\loadtable};
\addplot[mark = diamond*,mark options = {fill= white, thick}, white!00!blue, very thick] table [x index = 0, y index = 3] {\loadtable};
\addplot[mark = pentagon*, mark options = {fill= white, thick}, white!00!black, very thick] table [x index = 0, y index = 4] {\loadtable};
\addplot[mark = oplus*, mark options = {fill= white, thick}, white!00!violet, very thick] table [x index = 0, y index = 5] {\loadtable};
\addplot[mark = *, mark options = {fill= white, thick}, white!00!orange, very thick] table [x index = 0, y index = 6] {\loadtable};
\legend {$\mathcal{\hat{F}}_1$, $\mathcal{\hat{F}}_2$, $\mathcal{\hat{F}}_3$, $\mathcal{\hat{F}}_4$, $\mathcal{\hat{F}}_5$, $\mathcal{\hat{F}}_6$};
\end{axis}
\end{tikzpicture}
}
\caption{Impact of PSD.} \label{fig: NSF_power}
\end{figure}

First, we fix FEC OH at 7\% and compare different MFs as PSD varies. The simulation results of NSF network using 100 requests and 1,000~Gbps per request are illustrated in Fig.~\ref{fig: NSF_MF_power}.  It is observed in Fig.~\ref{fig: NSF_MF_power} that the revenue of different MFs increases as PSD changes from scenario 1 to 2, but then decreases from scenario 2 to 3. Both adaptive MF $\hat{\mathcal{M}}_3$ and $\hat{\mathcal{M}}_4$ that contain MFs $\{$BPSK,QPSK,8QAM$\}$ get the largest revenue in scenario 2, with 118\% improvement compared to $\hat{\mathcal{M}}_1$. It can be explained by the SNR threshold and spectral efficiency of different MFs. Only in the scenario with high SNR, high-order MFs can be adopted, which reduces the spectrum usage and spares more spectrum resources for other requests. However, in the scenario with low SNR, the adaptive method with four MFs has no difference with either one MF or two MFs, because the high-order MF cannot be adopted.


Then, we fix  MF at QPSK and compare multiple FECs. The results are illustrated in Fig. \ref{fig: NSF_FEC_power}, where we can see that, also in this case, as PSD changes from scenario 1 to 2, the revenue of different FEC increases, while from scenario 2 to 3, the revenue decreases. Different from the adaptive MF, having multiple FEC choices has a tiny impact on revenue difference on scenario 2, while a bigger difference is only observed for both scenarios 1 and 3, which is a different result with respect to  adaptive MF. In low SNR scenario, most lightpaths with small FEC OHs are blocked, 
while the redundant FEC with large FEC OHs can lower the SNR requirement and provide more SNR margins to overcome the PLIs. But in high SNR scenario, many requests have adopted the transmission mode $f_1$ with the highest spectral efficiency. Therefore, no revenue improvement can be observed in this scenario.



Figures \ref{fig: NSF_MF_power} and \ref{fig: NSF_FEC_power} indicate that adaptive MF brings more revenue in high SNR with median PSD, while multiple FECs configuration brings more revenue in low SNR scenarios. 
As we introduce more MFs and FEC, the high-order MF will mitigate the resource crunch and the low-spectral efficiency FEC with large OH can mitigate the PLIs.

\begin{figure*}[!htbp]
\centering
\subfigure[$G_i$=-18~dBm/GHz]{
\label{fig: nsf_high_snr_MmulFmul}
\scalebox{0.45}{
\begin{tikzpicture}
\begin{axis}[
width = 9 cm,
height = 7cm,
bar width=5pt,
ybar=5*\pgflinewidth,
ymin = 0, ymax = 230,
xmin = -0.5, xmax = 3.5,
ylabel = {Revenue},
ylabel near ticks,
xlabel = {Adaptive MF configuration},
xtick={0,1,2,3},
xticklabels = {$\hat{\mathcal{M}}_1$, $\hat{\mathcal{M}}_2$, $\hat{\mathcal{M}}_3$, $\hat{\mathcal{M}}_4$, $\hat{\mathcal{M}}_5$, $\hat{\mathcal{M}}_6$},
xmajorgrids,
ymajorgrids,
grid style = dashed,
legend pos = north west,
legend style ={fill =none,draw =none},
legend columns = 6,
       label style={font=\small},
       tick label style={font=\small},
]

\addplot[black!60!black, fill = white!20!green, postaction = {pattern = north west lines}
] table [x expr= \coordindex, y index = 1] {\loadedtableNSFM};

\addplot[black!60!black, fill=white!20!red, postaction = {pattern = north east lines}
] table [x expr= \coordindex, y index = 2] {\loadedtableNSFM};

\addplot[black!60!black, fill=white!20!blue, postaction = {pattern = north west lines}
] table [x expr= \coordindex, y index = 3] {\loadedtableNSFM};

\addplot[black!60!black, fill=white!20!purple, postaction = {pattern = north east lines}
] table [x expr= \coordindex, y index = 4] {\loadedtableNSFM};

\addplot[black!60!black, fill=white!20!cyan, postaction = {pattern = north west lines}
] table [x expr= \coordindex, y index = 5] {\loadedtableNSFM};

\addplot[black!00!black, fill=white!20!brown, postaction = {pattern = north east lines}
] table [x expr= \coordindex, y index = 6] {\loadedtableNSFM};
\legend{$\hat{\mathcal{F}}_1$,$\hat{\mathcal{F}}_2$,$\hat{\mathcal{F}}_3$,$\hat{\mathcal{F}}_4$,$\hat{\mathcal{F}}_5$,$\hat{\mathcal{F}}_6$};
\end{axis}

\end{tikzpicture}
}
}
\subfigure[$G_i$=-9~dBm/GHz]{
\label{fig: nsf_low_snr_MmulFmul}
\scalebox{0.45}{
\begin{tikzpicture}
\begin{axis}[
width = 9 cm,
height = 7cm,
bar width=5pt,
ybar=5*\pgflinewidth,
ymin = 0, ymax = 230,
ylabel = {Revenue},
ylabel near ticks,
xlabel = {Multiple FEC configuration},
xtick={0,1,2,3,4,5},
xticklabels={$\hat{\mathcal{F}}_1$,$\hat{\mathcal{F}}_2$,$\hat{\mathcal{F}}_3$,$\hat{\mathcal{F}}_4$,$\hat{\mathcal{F}}_5$,$\hat{\mathcal{F}}_6$},
xmajorgrids,
ymajorgrids,
grid style = dashed,
legend pos = north west,
legend style ={fill =none,draw =none},
legend columns = 5,
       label style={font=\small},
       tick label style={font=\small},
]

\addplot[black!60!black, fill = white!20!green, postaction = {pattern = north west lines}
] table [x expr= \coordindex, y index = 1] {\loadedtableNSFF};

\addplot[black!60!black, fill=white!20!red, postaction = {pattern = north east lines}
] table [x expr= \coordindex, y index = 2] {\loadedtableNSFF};

\addplot[black!60!black, fill=white!20!blue, postaction = {pattern = north west lines}
] table [x expr= \coordindex, y index = 3] {\loadedtableNSFF};

\addplot[black!60!black, fill=white!20!purple, postaction = {pattern = north east lines}
] table [x expr= \coordindex, y index = 4] {\loadedtableNSFF};
\legend{$\hat{\mathcal{M}}_1$, $\hat{\mathcal{M}}_2$, $\hat{\mathcal{M}}_3$, $\hat{\mathcal{M}}_4$};
\end{axis}

\end{tikzpicture}
}
}
\subfigure[$G_i$=-18~dBm/GHz]{
\label{fig: usb_high_snr_MmulFmul}
\scalebox{0.45}{
\begin{tikzpicture}
\begin{axis}[
width = 9 cm,
height = 7cm,
bar width=5pt,
ybar=5*\pgflinewidth,
ymin = 0, ymax = 230,
xmin = -0.5, xmax = 3.5,
ylabel = {Revenue},
ylabel near ticks,
xlabel = {Adaptive MF configuration},
xtick={0,1,2,3},
xticklabels = {$\hat{\mathcal{M}}_1$, $\hat{\mathcal{M}}_2$, $\hat{\mathcal{M}}_3$, $\hat{\mathcal{M}}_4$, $\hat{\mathcal{M}}_5$, $\hat{\mathcal{M}}_6$},
xmajorgrids,
ymajorgrids,
grid style = dashed,
legend pos = north west,
legend style ={fill =none,draw =none},
legend columns = 6,
       label style={font=\small},
       tick label style={font=\small},
]

\addplot[black!60!black, fill = white!00!green, postaction = {pattern = north west lines}
] table [x expr= \coordindex, y index = 1] {\loadedtableUSM};

\addplot[black!60!black, fill=white!20!red, postaction = {pattern = north east lines}
] table [x expr= \coordindex, y index = 2] {\loadedtableUSM};

\addplot[black!60!black, fill=white!20!blue, postaction = {pattern = north west lines}
] table [x expr= \coordindex, y index = 3] {\loadedtableUSM};

\addplot[black!60!black, fill=white!20!purple, postaction = {pattern = north east lines}
] table [x expr= \coordindex, y index = 4] {\loadedtableUSM};

\addplot[black!60!black, fill=white!20!cyan, postaction = {pattern = north west lines}
] table [x expr= \coordindex, y index = 5] {\loadedtableUSM};

\addplot[black!00!black, fill=white!20!brown, postaction = {pattern = north east lines}
] table [x expr= \coordindex, y index = 6] {\loadedtableUSM};
\legend{$\hat{\mathcal{F}}_1$,$\hat{\mathcal{F}}_2$,$\hat{\mathcal{F}}_3$,$\hat{\mathcal{F}}_4$,$\hat{\mathcal{F}}_5$,$\hat{\mathcal{F}}_6$};
\end{axis}

\end{tikzpicture}
}
}
\subfigure[$G_i$=-9~dBm/GHz]{
\label{fig: usb_low_snr_MmulFmul}
\scalebox{0.45}{
\begin{tikzpicture}
\begin{axis}[
width = 9 cm,
height = 7cm,
bar width=5pt,
ybar=5*\pgflinewidth,
ymin = 0, ymax = 230,
ylabel = {Revenue},
ylabel near ticks,
xlabel = {Multiple FEC configuration},
xtick={0,1,2,3,4,5},
xticklabels={$\hat{\mathcal{F}}_1$,$\hat{\mathcal{F}}_2$,$\hat{\mathcal{F}}_3$,$\hat{\mathcal{F}}_4$,$\hat{\mathcal{F}}_5$,$\hat{\mathcal{F}}_6$},
xmajorgrids,
ymajorgrids,
grid style = dashed,
legend pos = north west,
legend style ={fill =none,draw =none},
legend columns = 5,
       label style={font=\small},
       tick label style={font=\small},
]

\addplot[black!60!black, fill = white!20!green, postaction = {pattern = north west lines}
] table [x expr= \coordindex, y index = 1] {\loadedtableUSF};

\addplot[black!60!black, fill=white!20!red, postaction = {pattern = north east lines}
] table [x expr= \coordindex, y index = 2] {\loadedtableUSF};

\addplot[black!60!black, fill=white!20!blue, postaction = {pattern = north west lines}
] table [x expr= \coordindex, y index = 3] {\loadedtableUSF};

\addplot[black!60!black, fill=white!20!purple, postaction = {pattern = north east lines}
] table [x expr= \coordindex, y index = 4] {\loadedtableUSF};
\legend{$\hat{\mathcal{M}}_1$, $\hat{\mathcal{M}}_2$, $\hat{\mathcal{M}}_3$, $\hat{\mathcal{M}}_4$};
\end{axis}
\end{tikzpicture}
}
}
\caption{Revenue impact of joint MF and FEC. (a) and (b) are in NSF network; (c) and (d) are in US Backbone network.}
\label{fig: joint_mf_fec}
\end{figure*}
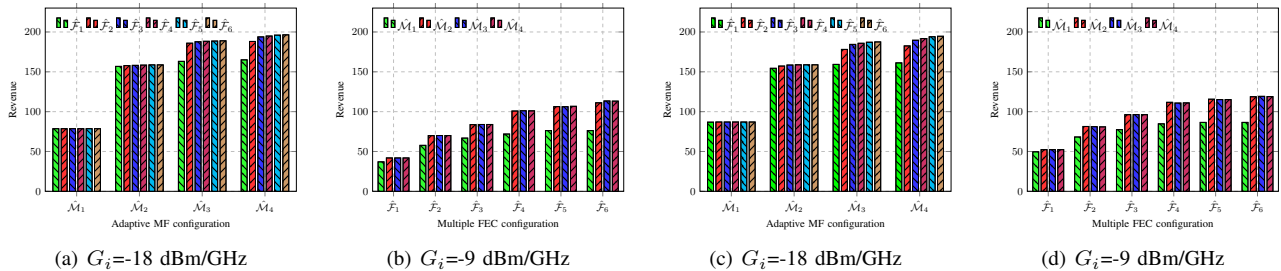

Given suitable PSD scenario of multiple FECs and adaptive MF, we further investigate the impact of joint MF and FEC schemes in NSF and US Backbone network. The revenues of 100 requests with average bit-rate 1,000 Gbps are illustrated in Fig. \ref{fig: joint_mf_fec}. In high SNR scenario with median PSD ($G_i=$-18~dBm/GHz), we find that the adaptive MFs enable to improve the revenue, while the configuration of multiple FECs has weak impact on the revenue. In low SNR scenario with high PSD ($G_i=$-9~dBm/GHz), both adaptive MFs and multiple FECs enable to improve the revenue, which means that the combination of MF and FEC is preferred in high PSD scenario rather than median PSD scenario. We also find that the revenue of adaptive MF configuration $\hat{\mathcal{M}}_3$ and multiple FEC configuration  $\hat{\mathcal{F}}_5$ can reach the almost maximum value for both high and low SNR scenarios. It means that the usage of MF with PM-16QAM and FEC OH with 50\% can be saved.


\pgfplotstableread{
F	M1	M2	M3	M4
F1	37.0	41.9	41.9	41.9
F2	57.7	69.9	70.0	70.0
F3	66.8	83.7	83.9	83.9
F4	71.9	100.8	101.1	101.0
F5	76.2	106.3	106.2	106.7
F6	76.2	111.1	113.4	113.2
}\loadedtableNSFF

\pgfplotstableread{
M	F1	F2	F3	F4	F5	F6
M1	78.6	78.6	78.6	78.6	78.6	78.6
M2	156.8	157.8	158.1	158.4	158.7	158.7
M3	163.1	185.9	187.7	188.3	188.6	188.9
M4	165.0	188.1	193.8	194.9	196.0	196.4
}\loadedtableNSFM

\pgfplotstableread{
F	M1	M2	M3	M4
F1	49.6 	52.3 	52.3 	52.3 
F2	68.2 	81.4 	81.1 	80.9 
F3	77.4 	96.2 	96.3 	96.2 
F4	84.7 	111.7 	110.8 	111.0 
F5	86.5 	115.6 	115.1 	115.1 
F6	86.5 	118.7 	119.2 	118.9 
}\loadedtableUSF

\pgfplotstableread{
M	F1	F2	F3	F4	F5	F6
M1	86.8 	87.0 	87.0 	87.0 	87.0 	87.0 
M2	154.4 	157.2 	158.5 	158.8 	158.8 	158.8 
M3	159.3 	178.1 	184.4 	185.8 	187.1 	187.5 
M4	161.2 	182.5 	189.6 	191.8 	194.0 	194.6 
}\loadedtableUSM

\subsection{Different traffic loads}

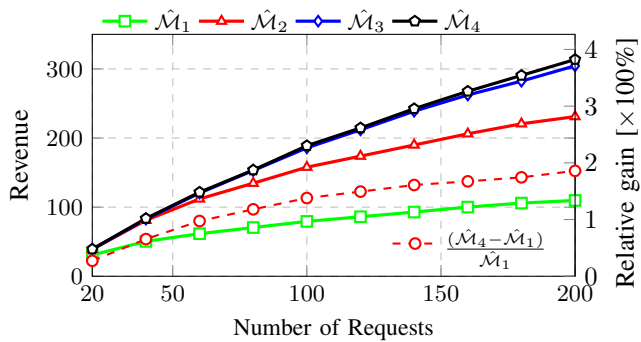
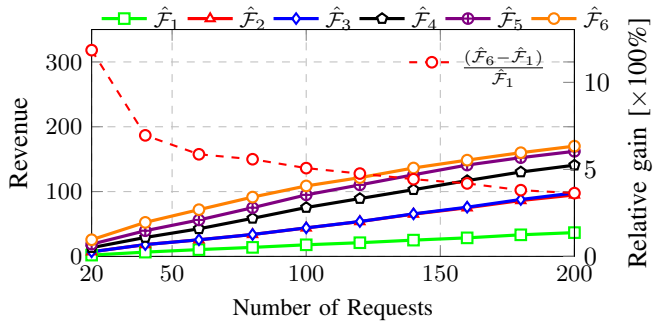
\begin{figure}[!htbp]
\centering
\subfigure[MF]{
\label{fig: NSF_MF_request}
\begin{tikzpicture}
\begin{axis}[
label style={font=\small},
tick label style={font=\small},
width = 8cm,
height = 4.8cm,
ymin = 0, ymax = 3.5e2,
xmin = 20, xmax = 200,
ylabel near ticks,
xtick = {20, 50, 100, 150, 200},
xlabel ={Number of Requests},
ylabel ={Revenue},
legend pos =north west,
legend columns = 4,
legend style={font =\footnotesize,draw=none,fill =none, at={(0,1.15)},anchor=north west},
xmajorgrids,
ymajorgrids,
grid style = dashed]
\addplot[mark=square*, mark options = {fill =white, thick}, white!0!green, very thick] table [x index=0, y index = 1] {\loadedtableNSFMFREQUEST};
\addplot[mark=triangle*, mark options = {fill = white, thick}, white!0!red,  very thick] table [x index=0, y index = 2]{\loadedtableNSFMFREQUEST};
\addplot[mark=diamond*, mark options = {fill = white, thick}, white!0!blue,  very thick] table [x index=0, y index = 3]{\loadedtableNSFMFREQUEST};
\addplot[mark=pentagon*, mark options = {fill = white, thick}, white!0!black,  very thick] table [x index=0, y index = 4]{\loadedtableNSFMFREQUEST};
\legend{$\mathcal{\hat{M}}_1$,$\mathcal{\hat{M}}_2$, $\mathcal{\hat{M}}_3$,$\mathcal{\hat{M}}_4$};
\end{axis}

\begin{axis}
[width = 8cm,
height = 4.6cm,
ymin = 0,   ymax = 4,
xmin = 20, xmax = 200,
axis y line* = right,
hide x axis,
ylabel={Relative gain [$\times$100\%]},
legend pos = south east,
legend style={font =\footnotesize,draw=none,fill =none},
]
\addplot[mark=*, mark options ={solid,fill=white}, dashed, thick, red] table [x index = 0, y expr = \thisrow{4} / \thisrow{1}-1]{\loadedtableNSFMFREQUEST};
\addlegendentry{$\frac{(\mathcal{\hat{M}}_4-\mathcal{\hat{M}}_1)}{\mathcal{\hat{M}}_1}$};

\end{axis}
\end{tikzpicture}
}
\subfigure[FEC]{
\label{fig: NSF_FEC_request}
\begin{tikzpicture}
\begin{axis}[
      label style={font=\small},
      tick label style={font=\small},
width = 8cm,
height = 4.6cm,
ymin = 0, ymax = 3.5e2,
xmin = 20, xmax = 200,
ylabel near ticks,
xtick = {20, 50, 100, 150, 200},
xlabel={Number of Requests},
ylabel={Revenue},
legend pos = north west,
legend style ={font=\footnotesize,draw=none,fill =none,at={(0,1.15)},anchor=north west,},
legend columns =6,
xmajorgrids,
ymajorgrids,
grid style =dashed
]

\addplot[mark = square*, mark options = {fill =white, thick}, white!00!green, very thick] table [x index = 0, y index = 1] {\loadedtableNSFFECREQUEST};
\addplot[mark = triangle*,mark options = {fill= white, thick}, white!00!red, very thick] table [x index = 0, y index = 2] {\loadedtableNSFFECREQUEST};
\addplot[mark = diamond*,mark options = {fill= white, thick}, white!00!blue, very thick] table [x index = 0, y index = 3] {\loadedtableNSFFECREQUEST};
\addplot[mark = pentagon*, mark options = {fill= white, thick}, white!00!black, very thick] table [x index = 0, y index = 4] {\loadedtableNSFFECREQUEST};
\addplot[mark = oplus*, mark options = {fill= white, thick}, white!00!violet, very thick] table [x index = 0, y index = 5] {\loadedtableNSFFECREQUEST};
\addplot[mark = *, mark options = {fill= white, thick}, white!00!orange, very thick] table [x index = 0, y index = 6] {\loadedtableNSFFECREQUEST};
\legend {$\mathcal{\hat{F}}_1$,$\mathcal{\hat{F}}_2$,$\mathcal{\hat{F}}_3$,$\mathcal{\hat{F}}_4$,$\mathcal{\hat{F}}_5$, $\mathcal{\hat{F}}_6$};
\end{axis}

\begin{axis}
[width = 8cm,
height = 4.6cm,
ymin = 0,  
xmin = 20, xmax = 200,
axis y line* = right,
hide x axis,
ylabel={Relative gain [$\times$100\%]},
legend pos = north east,
legend style={font =\footnotesize,draw=none,fill =none},
]
\addplot[mark=*, mark options ={solid,fill=white}, dashed, thick, red] table [x index = 0, y expr = \thisrow{6} / \thisrow{1}-1]{\loadedtableNSFFECREQUEST};
\addlegendentry{$\frac{(\mathcal{\hat{F}}_6-\mathcal{\hat{F}}_1)}{\mathcal{\hat{F}}_1}$};

\end{axis}

\end{tikzpicture}
}
\caption{Revenue impact with the number of requests in NSF network. Bit rate per request is 1,000 Gbps.}\label{fig: NSF_request}
\end{figure}

Let us now study the impact of different numbers of requests. The simulations assume all requests with identical 1,000 Gbps\cite{YADW17}. The PSD is either -18~dBm/GHz or -9~dBm/GHz,  such that we operate in scenarios that benefit of adaptive MFs and multiple FECs, respectively. The results of different adaptive MFs and different FECs are shown in Fig.~\ref{fig: NSF_MF_request} and \ref{fig: NSF_FEC_request}. In Fig. \ref{fig: NSF_MF_request}, the four adaptive MFs obtain the same result with 20 requests, but, as the number of requests increases, the gain achieved by using four different adaptive MFs also increases. The maximum improvement of adaptive MFs  (186\% higher compared to $\hat{\mathcal{M}}_1$) is obtained with 200 requests. In Fig. \ref{fig: NSF_FEC_request},  multiple FECs' revenue also increases with the number of requests. Configuration $\hat{\mathcal{F}}_6$ gets the largest revenue, which is 3.6 times higher than $\hat{\mathcal{F}}_1$.



\begin{figure}[!htbp]
\centering
\subfigure[MF]{
\label{fig: NSF_MF_traffic}
\begin{tikzpicture}
\begin{axis}[
       label style={font=\small},
       tick label style={font=\small},
width = 8cm,
height = 4.6cm,
ymin = 0, ymax = 3.5e2,
xmin = 250, xmax = 1500,
xtick = {250, 500, 750, 1000, 1250,1500},
ylabel near ticks,
xlabel ={Average bit rate [Gbps]},
ylabel ={Revenue},
legend pos =south west,
legend columns = 4,
legend style={font =\footnotesize,fill =none, draw=none,at={(0,1.15)},anchor =north west,},
xmajorgrids,
ymajorgrids,
grid style = dashed]

\addplot[mark=square*, mark options = {fill =white, thick}, white!0!green, very thick] table [x index=0, y index = 1] {\loadedtableNSFMFTRAFFIC};
\addplot[mark=triangle*, mark options = {fill = white, thick}, white!0!red,  very thick] table [x index=0, y index = 2]{\loadedtableNSFMFTRAFFIC};
\addplot[mark=diamond*, mark options = {fill = white, thick}, white!0!blue,  very thick] table [x index=0, y index = 3]{\loadedtableNSFMFTRAFFIC};
\addplot[mark=pentagon*, mark options = {fill = white, thick}, white!0!black,  very thick] table [x index=0, y index = 4]{\loadedtableNSFMFTRAFFIC};
\legend{$\mathcal{\hat{M}}_1$,$\mathcal{\hat{M}}_2$, $\mathcal{\hat{M}}_3$,$\mathcal{\hat{M}}_4$};
\end{axis}
\begin{axis}
[width = 8cm,
height = 4.6cm,
ymin = 0,   ymax = 4,
xmin = 250, xmax = 1500,
axis y line* = right,
hide x axis,
ylabel={Relative gain [$\times$100\%]},
legend pos = north east,
legend style={font =\footnotesize,draw=none,fill =none},
]
\addplot[mark=*, mark options ={solid,fill=white}, dashed, thick, red] table [x index = 0, y expr = \thisrow{4} / \thisrow{1}-1]{\loadedtableNSFMFTRAFFIC};
\addlegendentry{$\frac{(\mathcal{\hat{M}}_4-\mathcal{\hat{M}}_1)}{\mathcal{\hat{M}}_1}$};

\end{axis}
\end{tikzpicture}
}
\subfigure[FEC]{
\label{fig: NSF_FEC_traffic}
\begin{tikzpicture}
\begin{axis}[
   label style={font=\small},
   tick label style={font=\small},
width = 8cm,
height = 4.6cm,
ymin = 0, ymax = 3.5e2,
xmin = 250, xmax = 1500,
xtick = {250, 500, 750, 1000, 1250,1500},
ylabel near ticks,
xlabel={Average bit rate [Gbps]},
ylabel={Revenue},
legend pos = north west,
legend style ={font=\footnotesize,fill=none, draw=none, at={(0,1.15)},anchor = north west},
legend columns =6,
xmajorgrids,
ymajorgrids,
grid style =dashed
]
\addplot[mark = square*, mark options = {fill =white, thick}, white!00!green, very thick] table [x index = 0, y index = 1] {\loadedtableNSFFECTRAFFIC};
\addplot[mark = triangle*,mark options = {fill= white, thick}, white!00!red, very thick] table [x index = 0, y index = 2] {\loadedtableNSFFECTRAFFIC};
\addplot[mark = diamond*,mark options = {fill= white, thick}, white!00!blue, very thick] table [x index = 0, y index = 3] {\loadedtableNSFFECTRAFFIC};
\addplot[mark = pentagon*, mark options = {fill= white, thick}, white!00!black, very thick] table [x index = 0, y index = 4] {\loadedtableNSFFECTRAFFIC};
\addplot[mark = oplus*, mark options = {fill= white, thick}, white!00!violet, very thick] table [x index = 0, y index = 5] {\loadedtableNSFFECTRAFFIC};
\addplot[mark = *, mark options = {fill= white, thick}, white!00!orange, very thick] table [x index = 0, y index = 6] {\loadedtableNSFFECTRAFFIC};

\legend {$\mathcal{\hat{F}}_1$,$\mathcal{\hat{F}}_2$,$\mathcal{\hat{F}}_3$,$\mathcal{\hat{F}}_4$,$\mathcal{\hat{F}}_5$, $\mathcal{\hat{F}}_6$};
\end{axis}
\begin{axis}
[width = 8cm,
height = 4.6cm,
ymin = 0,  ymax = 4,
xmin = 250, xmax = 1500,
axis y line* = right,
hide x axis,
ylabel={Relative gain [$\times$100\%]},
legend pos = south east,
legend style={font =\footnotesize,draw=none,fill =none, at={(0.65,0.5)},anchor =south west},
]
\addplot[mark=*, mark options ={solid,fill=white}, dashed, thick, red] table [x index = 0, y expr = \thisrow{6} / \thisrow{1}-1]{\loadedtableNSFFECTRAFFIC};
\addlegendentry{$\frac{(\mathcal{\hat{F}}_6-\mathcal{\hat{F}}_1)}{\mathcal{\hat{F}}_1}$};

\end{axis}
\end{tikzpicture}
}
\caption{Revenue impact with different traffic rates in NSF network. The simulations use 160 requests.}\label{fig: NSF_traffic}
\end{figure}
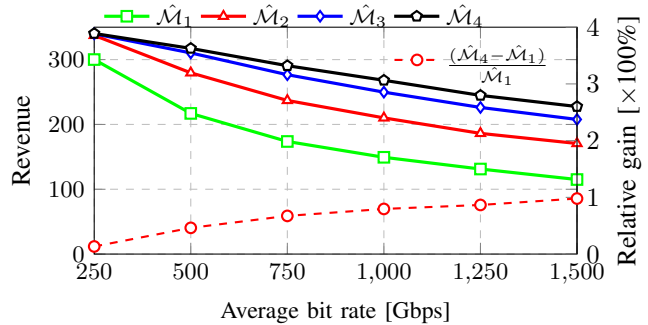
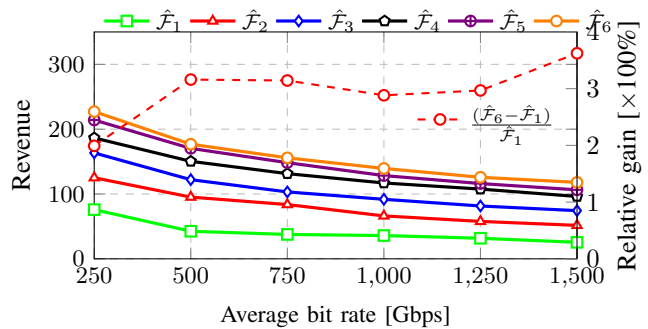

We report the simulation results with different average bit-rates in Fig. \ref{fig: NSF_traffic}. 
For a given average bit rate of 250+$n$*250, each request can randomly chose the bit-rate from the set $\{$250, ..., 250+2$n$*250$\}$. 160 requests are assumed in the simulation. The results of different MFs and FECs are shown in Figs. \ref{fig: NSF_MF_traffic} and \ref{fig: NSF_FEC_traffic}. We observe that the revenue decreases with the average bit-rate. The larger the bit-rate, the more spectral resources' consumption of fiber, which leads to blocked requests. For the case with more transmission modes, such as $\hat{\mathcal{M}}_4$ or $\hat{\mathcal{F}}_6$, it can also gain more revenue compared to the other configuration with fewer transmission modes, $\hat{\mathcal{M}}_1$ or $\hat{\mathcal{F}}_1$. When the average bit-rate increases to 1,500~Gbps, the revenue improvement ratio of $\hat{\mathcal{M}}_4$ and $\hat{\mathcal{F}}_6$ reaches about 98\% and 362\% compared to $\hat{\mathcal{M}}_1$ and $\hat{\mathcal{F}}_1$, respectively.


\pgfplotstableread{
20	30.8	39	39.2	39.2
40	50.5	81	82.7	83.6
60	61.6	111.9	120.4	121.6
80	70.5	134.6	153	153.7
100	79.4	157.8	185.8	188.9
120	86.1	173.8	211.9	214.6
140	92.9	189.9	238.8	242.4
160	100.1	206.1	262.3	267.6
180	105.9	220.6	282.3	290.6
200	109.7	231	304.5	313.6
}\loadedtableNSFMFREQUEST

\pgfplotstableread{
250	300.1	337.4	339.8	340.3
500	217	279.6	310.3	317.3
750	173.7	237	276.7	290.6
1000	149.3	210.1	249.8	268.1
1250	131.2	186.2	226.2	244.8
1500	115	170.6	207.7	227.6
}\loadedtableNSFMFTRAFFIC

\pgfplotstableread{
20	2	7.1	7.1	13.5	18.8	25.7
40	6.6	18.2	18.2	29.2	39.6	52.5
60	10.5	25.6	25.6	42.6	56.2	72.1
80	13.9	33.8	33.8	58.7	75.1	91.5
100	17.9	43.9	44	75.4	94.9	108.7
120	21.1	53.7	53.8	89.2	110.3	121.5
140	25.1	65.2	65.6	102.9	125.9	136.4
160	28.6	75	75.9	116.9	140.9	148.5
180	33.3	86.2	87.8	130.2	152.3	160
200	36.7	94.8	97.9	140.7	162.1	170.1
}\loadedtableNSFFECREQUEST

\pgfplotstableread{
250	75.9	125.1	163.4	186.5	214.1	227.1
500	42.5	95.2	122.1	150.4	170.4	176.8
750	37.6	83.7	103.2	131.4	148.3	155.8
1000	35.9	66.2	91.8	117	128.4	139.4
1250	31.7	57.7	81.5	107.6	116.1	125.8
1500	25.5	51.6	74.3	96.2	106.4	117.9
}\loadedtableNSFFECTRAFFIC

\section{Conclusion}\label{sec: conclusions}

In this paper, we studied the problem of using adaptive MFs and multiple FECs to improve the traffic provisioning in FONs. The objective is to maximize the total network revenue. To this end, we develop a MILP model and a fast two-phase heuristic algorithm, which is shown to be near-optimal for revenue maximization. Although the revenue loss is inevitable under different
resource crunch scenarios, it can be improved by properly choosing the transmission mode configurations and physical parameters. Through simulations, we demonstrate that using adaptive MF enables to increase the revenue more than 100\% in the scenario of high SNR while using adaptive FEC is profitable for scenarios with low SNR. While guaranteeing the revenue performance, the usage of adaptive MF configuration with PM-16QAM and multiple FEC configuration with OH 50\% can be saved. We also carry out experiments to demonstrate the case of severe resource crunch,  which is simulated by increasing bit-rate and number of requests. It shows that for the case of high traffic load (large number of requests or big average bit-rate), adaptive MF takes more advantage than single MF with PM-BPSK, because it can offer more spectrum-efficient transmission modes.

\section*{Acknowledgment}
The work is jointly supported by Eiffel Scholarship (No. 895145D), open project (2020GZKF017) of Shanghai Jiao Tong University. China Scholarship Council (No. 201806230093), National Nature Science Fund of China (No.61775137, No.62071295, No.61431009, and No.61433009), National ``863'' Hi-tech Project of China (No.2013AA013602 and No.2012AA011301), and NSF (Grant No. 1716945).

\bibliographystyle{IEEEtran}
\bibliography{references}
\end{document}